\documentclass[useAMS,usenatbib]{mn2e}
\usepackage[pdftex]{graphicx}
\usepackage{float}
\usepackage{multirow}
\usepackage{amssymb}
\usepackage{amsmath}
\usepackage{caption}
\usepackage{mathptmx}
\usepackage{bm}
\usepackage{enumitem}  
\usepackage{booktabs} 
\usepackage{xspace}
\usepackage{pdflscape}
\usepackage{enumitem}
\usepackage{longtable} 
\usepackage{cases}
\usepackage{threeparttable}
\usepackage{threeparttablex}
\usepackage{todonotes}
		\newcommand{\msun}{\mbox{\,$M_{\odot}$}\xspace}

		\newcommand{\vunit}{\mbox{\,km\,s$^{-1}$}\xspace}

		\newcommand{\fexxv}{Fe\,\textsc{xxv}\xspace}
		
		\newcommand{\fexxvi}{Fe\,\textsc{xxvi}\xspace}

		\newcommand{\na}{--}
		\newcommand{\vout}{v_{\rm w}\xspace}
		\newcommand{\mout}{\dot M_{\rm w}}

		\newcommand{\nh}{N_{\rm H}}
		\newcommand{\lognh}{\log(N_{\rm H}/\rm{cm}^{-2})}
		\newcommand{\logxi}{\log(\xi/\rm{erg\,cm\,s}^{-1})}
		
		\newcommand{\hea}{He$\alpha$\xspace}
		\newcommand{\lya}{Ly$\alpha$\xspace}
		\newcommand{\lbol}{L_{\rm bol}} 
		\newcommand{\lk}{L_{\rm w}}
		\newcommand{\lkmax}{\lk^{\rm max}}
		\newcommand{\lkmin}{\lk^{\rm min}}
		
		\newcommand{\moutmax}{\mout^{\rm max}}
		\newcommand{\moutmin}{\mout^{\rm min}}
		\newcommand{\rmin}{r_{\rm min}}
		\newcommand{\resc}{r_{\rm esc}}
		\newcommand{\rmax}{r_{\rm max}}
		\newcommand{\eddratio}{\lambda}
		\newcommand{\lcorr}{\kappa_{\rm bol}}
		\newcommand{\lion}{L_{\rm ion}}
		\newcommand{\ledd}{L_{\rm Edd}}
		\newcommand{\medd}{\dot M_{\rm Edd}}
		\newcommand{\mbh}{M_{\rm BH}}
		\newcommand{\pout}{\dot p_{\rm w}}
		\newcommand{\poutmax}{\pout^{\rm max}}
		\newcommand{\poutmin}{\pout^{\rm min}}
		\newcommand{\pedd}{\dot p_{\rm Edd}}
		\newcommand{\vesc}{v_{\rm esc}}
		\newcommand{\rs}{r_{\rm s}}
		\newcommand{\tesc}{\tau}
		\newcommand{\msigma}{\mbh-\sigma_{\ast}} 
		\def\pbol{\dot p_{\rm bol}}

		\newcommand{\suzaku}{\emph{Suzaku}\xspace}
		\newcommand{\xmm}{\emph{XMM-Newton}\xspace}

		\newcommand{\xstar}{\textsc{xstar}\xspace}

\bibpunct{[}{]}{,}{a}{,}{;}
\title[The Suzaku view of highly-ionised outflows in AGN, Paper II]{The \suzaku view of highly-ionised outflows in AGN: II -- Location, energetics and scalings with Bolometric Luminosity.}
\author[J.~Gofford et al.]{J.~Gofford$^{1,2}$\thanks{E-mail: jgofford@umbc.edu}, J.~N.~Reeves$^{1,2}$, D.~E.~McLaughlin$^{2}$, V.~Braito$^{3}$, T.~J.Turner$^{1}$, \newauthor F.~Tombesi$^{4,5}$,  M.~Cappi$^{6}$
\\
$^{1}$Department of Physics, University of Maryland Baltimore County, Baltimore, MD 21250, USA \\
$^{2}$Astrophysics Group, Keele University, Keele, ST5 5BG, UK \\
$^{3}$INAF-Osservatorio Astronomico di Brera, Via Bianchi 46, I-23807 Merate, Italy \\
$^{4}$X-ray Astrophysics Laboratory and CRESST, NASA/GSFC, Greenbelt, MD 20771, USA \\
$^{5}$Department of Astronomy, University of Maryland, College Park, MD 20742, USA \\
$^{6}$INAF-IASF Bologna, Via Gobetti 101, I-40129, Bologna, Italy \\
}

\begin{document}

\date{\today}

\pagerange{\pageref{}--\pageref{}} \pubyear{?}

\maketitle
\label{firstpage}   
\begin{abstract} 
Ongoing studies with \xmm have shown that powerful accretion disc winds, as revealed through highly-ionised Fe\,K-shell absorption at $E\geq6.7$\,keV, are present in a significant fraction of Active Galactic Nuclei (AGN) in the local Universe (\citealt{tombesi2010a}). In \citet{gofford2013} we analysed a sample of 51 \suzaku-observed AGN and independently detected Fe\,K absorption in $\sim40\%$ of the sample, and we measured the properties of the absorbing gas. In this work we build upon these results to consider the properties of the associated wind. On average, the fast winds ($\vout>0.01$\,c) are located $\langle r \rangle\sim10^{15-18}$\,cm (typically $\sim10^{2-4}\,\rs$) from their black hole, their mass outflow rates are of the order $\langle \mout \rangle\sim0.01-1$\,\msun\,yr$^{-1}$ or $\sim(0.01-1)\medd$ and kinetic power is constrained to $\langle \lk \rangle\sim10^{43-45}$\,erg\,s$^{-1}$, equivalent to $\sim(0.1-10\%)\ledd$. We find a fundamental correlation between the source bolometric luminosity and the wind velocity, with $\vout\propto\lbol^{\alpha}$ and $\alpha=0.4^{+0.3}_{-0.2}$ ($90\%$ confidence), which indicates that more luminous AGN tend to harbour faster Fe\,K winds. The mass outflow rate $\mout$, kinetic power $\lk$ and momentum flux $\pout$ of the winds are also consequently correlated with $\lbol$, such that more massive and more energetic winds are present in more luminous AGN. We investigate these properties in the framework of a continuum-driven wind, showing that the observed relationships are broadly consistent with a wind being accelerated by continuum-scattering. We find that, globally, a significant fraction ($\sim85\%$) of the sample can plausibly exceed the $\lk/\lbol\sim0.5\%$ threshold thought necessary for feedback, while $45\%$ may also exceed the less conservative $\sim5\%$ of $\lbol$threshold as well. This suggests that the winds may be energetically significant for AGN--host-galaxy feedback processes. 
\end{abstract}

\section{introduction}
\label{sec:introduction}
Outflows of photo-ionised gas are ubiquitous in Active Galactic Nuclei (AGN). Recent surveys with \xmm and \suzaku by \citet{tombesi2010b} and \citet{gofford2013} (hereafter `Paper~I') have shown that Fe\,K-shell outflows are observed in a significant fraction ($\sim40-50\%$) of active galaxies. These potentially massive outflows ($\nh\sim10^{23}$\,cm$^{-2}$), which are identified through blue-shifted resonant absorption from \fexxv~\hea ($E_{\rm rest}=6.7$\,keV) and \fexxvi~\lya ($E_{\rm rest}=6.97$\,keV), can posses outflow velocities ($\vout$) ranging from a few thousand km\,s$^{-1}$ up to mildly relativistic values ($\vout\sim0.1-0.3$\,c; \citealt{pounds2003, chartas2002, reeves2009, tombesi2010a, gofford2013,tombesi2015nature}) which indicates a substantial mass transport into the host galaxy. The large wind velocity --- which indicates an origin directly associated with the accretion disk, hence leading them to be dubbed `disk-winds' --- implies that the ensuing outflow may be energetically significant in terms of feedback (i.e., $L/\lbol\gtrsim0.5-5\%$, \citealt{hopkins2010,dimatteo2005}). The winds are observed in both radio-quiet (\citealt{tombesi2010a}, Paper~I) and radio-loud (Paper~I, \citealt{tombesi2014}) AGN which suggests that they may also be an important addition to the traditional AGN unified model (e.g., \citealt{antonucci1993,urry&padovani1995}). 

In Paper~I we performed a systematic search for Fe\,K absorption in 51 \suzaku-observed AGN which were heterogeneously selected from the {\sc heasarc} data archive\footnote{Accessible at: http://heasarc.gsfc.nasa.gov/docs/archive.html} (see Paper~I for selection criteria). Our main results from that paper were: (i) blue-shifted Fe\,K-shell absorption is present at the $\geq2\sigma$ level in 20/51 ($\sim40\%$) of the selected AGN (in 28/73 individual observations); (ii) the absorbing gas is typified by mean values of $\lognh\sim23$ and $\logxi\sim4.5$, while the outflow velocity spans a continuous range between $\vout<1,500$\,km\,s$^{-1}$ and $\lesssim100,000$\,km\,s$^{-1}$, respectively, with a mean and median velocity of $0.1$\,c and $0.056$\,c. Ultimately, these results are all consistent with those found by \citet{tombesi2010a} on the basis of their \xmm sample. 

In this work, we build upon our previous results to assess the properties of the disk-wind itself. The remainder of this paper is laid out as follows: in \S\ref{sec:Wind parameters} we outline the equations through which assess the wind properties, in \S\ref{sec:parameter_constraints} we use the inferred values to perform a qualitative assessment of the wind properties, before going on to conduct a quantitative correlation analysis of how the various wind properties scale with the AGN bolometric luminosity in \S\ref{sec:correlation_analyses}. In \S\ref{sec:discussion} we then compare our results to those already present in the literature, discuss the wind acceleration mechanism which may be responsible for the observed outflows, and finally assess whether the observed wind is likely to be energetically significant in terms of feedback. We summarise our results and present our overall conclusions in \S\ref{sec:summary_conclusions}.

\begin{table*}
	\begin{center}
	\begin{minipage}{12cm}
	\begin{threeparttable}
	\caption{Summary of AGN and wind properties\label{table:measured_wind_properties} }
	  \begin{tabular}{l c c c c c}
	  \toprule 
	  \multicolumn{1}{c}{\multirow{2}{*}{Source}} &
	  \multicolumn{1}{c}{$\log \mbh$} & 
	  \multicolumn{1}{c}{$\log \lion$}  & 
	  \multicolumn{1}{c}{$\log\nh$} &
	  \multicolumn{1}{c}{$\log\xi$} &
	  \multicolumn{1}{c}{$\vout/c$} 
	  \\[0.5ex]

	   & 
	  \multicolumn{1}{c}{(1)} & 
	  \multicolumn{1}{c}{(2)} & 
	  \multicolumn{1}{c}{(3)} &
	  \multicolumn{1}{c}{(4)} &
	  \multicolumn{1}{c}{(5)} 
	  \\
	  \midrule
	
	  3C\,111$^{\,3,4,\ast}$ 
	  	& $8.1\pm0.5$ & 44.7 & $23.0^{+0.4}_{-0.8}$ & $4.45^{+0.40}_{-0.46}$ & $0.072^{+0.041}_{-0.038}$ \\[0.5ex]

	  3C\,390.3$^{\,1,5}$ 
	  	& $8.8^{+0.2}_{-0.6}$ & 44.7 & $>23.7$ & $>5.46$ & $0.145\pm0.007$ \\[0.5ex]

	  4C\,+74.26$^{\,16}$ 
	  	& $9.6^{\dag}$ & 47.0 & $>21.8$ & $4.06\pm0.45$ & $0.185\pm0.026$ \\[0.5ex]

	  APM\,08279$^{\,17,a}$ 
	  	& $10.0^{\dag}$ & 45.4 & $23.0\pm0.1$ & $3.51^{+0.33}_{-0.18}$ & $0.285^{+0.165}_{-0.158}$ \\[0.5ex]

	  CBS\,126$^{\,15}$ 
	  	& $7.8\pm0.1$ & 44.2 & $>23.7$ & $4.77^{+0.26}_{-0.17}$ & $0.012\pm0.006$ \\[0.5ex]

	  ESO\,103-G035$^{\,14,\ast}$ 
	  	& $7.4\pm0.1$ & 43.9 & $>21.9$ & $4.36\pm1.19$ & $0.056\pm0.025$ \\[0.5ex]

	  MCG\,-6-30-15$^{\,7}$ 
	  	& $6.7\pm0.2$ & 43.2 & $22.2\pm0.1$ & $3.64\pm0.06$ & $0.007\pm0.002$ \\[0.5ex]

	  MR\,2251-178$^{\,6}$ 
	  	& $8.7\pm0.1$ & 45.2 & $21.5\pm0.2$ & $3.26\pm0.12$ & $0.137\pm0.008$ \\[0.5ex]

	  Mrk\,279$^{\,5}$ 
	  	& $7.5\pm0.1$ & 43.4 & $23.4\pm0.3$ & $4.42^{+0.15}_{-0.27}$ & $0.220\pm0.006$ \\[0.5ex]

	  Mrk\,766$^{\,8}$ 
	  	& $6.2^{+0.3}_{-0.6}$ & 43.2 & $22.7^{+0.2}_{-0.3}$ & $3.86^{+0.37}_{-0.25}$ & $0.039^{+0.030}_{-0.026}$ \\[0.5ex]

	  NGC\,1365$^{\,13,\ast}$ 
	  	& $7.6\pm0.7$ & 42.8 & $23.7^{+0.2}_{-0.5}$ & $3.88\pm0.07$ & $<0.014$ \\[0.5ex]

	  NGC\,3227$^{\,1}$ 
	  	& $7.6\pm0.2$ & 42.5 & $22.7^{+0.2}_{-0.3}$ & $3.88^{+0.12}_{-0.16}$ & $<0.008$ \\[0.5ex]

	  NGC\,3516$^{\,1}$ 
	  	& $7.6\pm0.1$ & 43.6 & $22.6\pm0.2$ & $3.84\pm0.11$ & $0.004\pm0.002$ \\[0.5ex]

	  NGC\,3783$^{\,1}$ 
	  	& $7.5\pm0.1$ & 43.6 & $21.8\pm0.2$ & $3.48^{+0.15}_{-0.07}$ & $<0.008$ \\[0.5ex]

	  NGC\,4051$^{\,1}$ 
	  	& $6.3\pm0.2$ & 42.5 & $22.8\pm0.1$ & $3.50^{+0.53}_{-0.50}$ & $0.018^{+0.004}_{-0.005}$ \\[0.5ex]

	  NGC\,4151$^{\,1}$ 
	  	& $7.1\pm0.1$ & 42.9 & $>21.7$ & $3.69\pm0.64$ & $0.055\pm0.023$ \\[0.5ex]

	  NGC\,4395$^{\,2}$ 
	  	& $4.7\pm0.2$ & 40.7 & $22.8\pm0.3$ & $3.92\pm0.16$ & $<0.001$ \\[0.5ex]

	  NGC\,5506$^{\,6,9,10,\ast}$ 
	  	& $7.3\pm0.7$ & 43.7 & $23.2\pm0.3$ & $5.04^{+0.29}_{-0.17}$ & $0.246\pm0.006$ \\[0.5ex]

	  PDS\,456$^{\,11,\ast}$ 
	  	& $9.4\pm0.3$ & 45.3 & $23.0\pm0.1$ & $4.06^{+0.28}_{-0.15}$ & $0.273\pm0.028$ \\[0.5ex]

	  SW\,J2127$^{\,12,b}$ 
		& $7.2^{\dag}$ & 43.7 & $22.8\pm0.3$ & $4.16^{+0.29}_{-0.13}$ & $0.231\pm0.006$ \\

	  \bottomrule
	  \end{tabular}
		\begin{tablenotes}
		\item[]\textsc{Columns:}-- (1) SMBH mass, in units of $\msun$; (2) absorption-corrected 1--1000 Rydberg ionising luminosity (in units of erg\,s$^{-1}$), as determined from the best-fit continuum model outlined in Paper I; (3) measured wind column density, in units of cm$^{-2}$; (4) wind ionisation parameter, in units of erg\,cm\,s$^{-1}$; (5) inferred wind outflow velocity, as measured in Paper~I.\\[-5pt]
		\item[]\textsc{References:}-- $^{1}$\citet{peterson2004}; $^{2}$\citet{edri2012}; $^{3}$\citet{chatterjee2011}; $^{4}$\citet{tombesi2012b}; $^{5}$\citet{bentz2009a}; $^{6}$\citet{khorunzhev2012}; $^{7}$\citet{mchardy2005}; $^{8}$\citet{bentz2009b}; $^{9}$\citet{papidakis2004}; $^{10}$\citet{nikolajuk2009}; $^{11}$\citet{reeves2009}; $^{12}$\citet{malizia2008}; $^{13}$\citet{risaliti2007}; $^{14}$\citet{czerny2001}; $^{15}$\citet{grupe2004}; $^{16}$\citet{woo2002}; $^{17}$\citet{saez2009}.\\[-5pt]
		\item[]\textsc{Notes:}-- $^{\ast}\mbh$ is taken as the mean of extreme values found in the literature; $^{\dag}$Errors on $\mbh$ not present in literature; $^{a}$Full designation APM\,08279+5255; $^{b}$Full designation SWIFT\,J2127.4+5654. Where a source has an Fe\,K wind detected in more than one observation, the mean $\nh$, $\xi$ and $\vout$ values are reported here; see Paper I for further details.
		\end{tablenotes}
 	\end{threeparttable}
	\end{minipage}
	\end{center}
\end{table*}

\section{Wind parameters}
\label{sec:Wind parameters}

\subsection{Projected distance}
\label{sub:projected_distance}
The maximum column density $\nh$ of gas along the line-of-sight (LOS) is given by $\nh=\int_{\rmax}^{\infty}n(r)~dr$, where $n(r)$ is the average gas number density and $\rmax$ is the maximum observed distance of the absorber. This, combined with the definition of the ionisation parameter $\xi=\lion/nr^{2}$ (\citealt{tarter1969}), yields the maximum distance that the absorber can be located from the ionising source given its observed column density and ionisation state,
\begin{equation}
	\rmax=\dfrac{\lion}{\xi\nh},
	\label{eq:rmax}
\end{equation}
where $\lion$ is the source ionising luminosity integrated between $1-1000$ Rydberg. Conversely, a lower limit on $r$ can be inferred by considering the escape radius of the gas given its observed velocity. For a simple Keplerian disc orbiting a black hole the escape velocity at distance $r$ is $\vesc=\sqrt{2G\mbh/r}$. In the limit that $\vout=\vesc$, i.e., assuming that the measured outflow velocity along the LOS is equal to the escape velocity at observed radius $r$, we can set a lower limit on the location of the wind,
\begin{equation}
	\rmin=\dfrac{2G\mbh}{\vout^{2}}.
	\label{eq:rmin}
\end{equation} 
These relations hence allow upper/lower limits to be placed on the wind location given the measured parameters of the absorbing gas, albeit with large uncertainties.

\begin{table*}
	\begin{center}
	\begin{minipage}{12cm}
	\begin{threeparttable}
	  \caption{Summary of inferred wind parameters.\label{table:main_results}}
	\footnotesize
	  \begin{tabular}{l c c c c c c}
	  \toprule
 
	  \multicolumn{1}{c}{\multirow{2}{*}{Source}} &
	  \multicolumn{1}{c}{$\log \rmin$}  & 
	  \multicolumn{1}{c}{$\log \rmax$}  & 
	  \multicolumn{1}{c}{$\log \moutmin$}  & 
	  \multicolumn{1}{c}{$\log \moutmax$} & 
	  \multicolumn{1}{c}{$\log \lkmin$}  & 
	  \multicolumn{1}{c}{$\log \lkmax$} \\[0.5ex]
	
	   & 
	  \multicolumn{1}{c}{(cm)}  &  
	  \multicolumn{1}{c}{(cm)}  & 
	  \multicolumn{1}{c}{$({\rm g\,s}^{-1})$}  & 
	  \multicolumn{1}{c}{$({\rm g\,s}^{-1})$}  & 
	  \multicolumn{1}{c}{$({\rm erg\,s}^{-1})$}  & 
	  \multicolumn{1}{c}{$({\rm erg\,s}^{-1})$} \\
	  \midrule
	
	  3C\,111 		
		& $15.9\pm0.1$ & $17.3\pm0.6$ & $24.9\pm0.1$ & $26.4\pm0.6$ & $43.3\pm0.1$ 
		& $44.7\pm0.6$ \\[0.5ex]

	  3C\,390.3 		
		& $15.7\pm0.1$ & $<16.7$ & $>25.5$ & $<26.6$ & $>44.5$ & $<45.6$ \\[0.5ex]

	  4C\,+74.26  	
		& $16.5\pm0.1$ & $<20.0$ & $>24.8$ & $27.9\pm0.5$ & $>44.0$ 
		& $47.1^{+0.5}_{-0.7}$\\[0.5ex]

	  APM\,08279  
		& $16.6\pm0.1$ & $19.8\pm0.3$ & $26.3\pm0.1$ & $29.6\pm0.3$ & $45.9\pm0.1$ 
		& $49.1\pm0.3$\\[0.5ex]

	  CBS\,126  
		& \na & $<15.9$ & \na & $24.8\pm0.3$ & \na & $41.6\pm0.2$ \\[0.5ex]

	  ESO\,103-G035  	
		& $15.4^{+0.5}_{-0.3}$ & $<18.8$ & $>23.0$ & $25.6\pm1.2$ & $>41.1$ 
		& $43.7\pm1.2$\\[0.5ex]

	  MCG\,-6-30-15  	
		& $16.5\pm0.3$ & $17.4\pm0.1$ & $23.8\pm0.3$ & $24.7\pm0.1$ & $40.1\pm0.3$ 
		& $41.0\pm0.1$\\[0.5ex]

	  MR\,2251-178  	
		& $15.9\pm0.1$ & $20.4\pm0.3$ & $24.3\pm0.1$ & $27.8\pm0.7$ & $43.3\pm0.1$ 
		& $46.7\pm0.7$\\[0.5ex]

	  Mrk\,279  		
		& $14.3\pm0.1$ & $15.6\pm0.5$ & $24.3\pm0.1$ & $25.6\pm0.5$ & $43.6\pm0.1$ 
		& $44.9\pm0.5$ \\[0.5ex]

	  Mrk\,766 		
		& $14.5\pm0.1$ & $16.7\pm0.3$ & $23.1\pm0.1$ & $25.2\pm0.3$ & $40.9\pm0.1$ 
		& $43.1\pm0.3$ \\[0.5ex]

	  NGC\,1365 		
		& \na & $15.2\pm0.1$ & \na & $24.1\pm0.1$ & \na & $40.5\pm0.1$ \\[0.5ex]

	  NGC\,3227 		
		& \na & $16.0\pm0.4$ & \na & $23.6\pm0.4$ & \na & $39.7\pm0.4$ \\[0.5ex]

	  NGC\,3516 		
		& \na & $17.2\pm0.3$ & \na & $24.6\pm0.3$ & \na & $40.5\pm0.3$ \\[0.5ex]

	  NGC\,3783  		
		& $>17.3$ & $18.4\pm0.3$ & $>24.2$ & $<25.5$ & $<40.5$ & $<41.9$ \\[0.5ex]

	  NGC\,4051 		
		& $15.3\pm0.1$ & $15.7\pm0.2$ & $23.6\pm0.1$ & $24.0\pm0.2$ & $40.7\pm0.1$ 
		& $41.2\pm0.2$ \\[0.5ex]

	  NGC\,4151 		
		& $15.1\pm0.4$ & $<18.1$ & $>22.6$ & $25.2\pm0.6$ & $>40.7$ & $43.4\pm0.6$ \\[0.5ex]

	  NGC\,4395 		
		& \na & $13.9\pm0.4$ & \na & $<21.5$ & \na & $<36.1$ \\[0.5ex]

	  NGC\,5506 		
		& $14.0\pm0.1$ & $15.4\pm0.5$ & $23.9\pm0.1$ & $25.3\pm0.5$ & $43.3\pm0.1$ 
		& $44.7\pm0.5$ \\[0.5ex]

	  PDS\,456 		
		& $16.0\pm0.1$ & $18.2\pm0.2$ & $25.8\pm0.1$ & $27.9\pm0.2$ & $45.3\pm0.1$ 
		& $47.5\pm0.2$ \\[0.5ex]

	  SW\,J2127 		
		& $13.9\pm0.1$ & $16.8\pm0.5$ & $23.4\pm0.1$ & $26.2\pm0.5$ & $42.8\pm0.1$ 
		& $45.6\pm0.5$ \\

		\bottomrule
		\end{tabular}
			\begin{tablenotes}
				\item[]
			\end{tablenotes}
	\end{threeparttable}
	\end{minipage}
	\end{center}
\end{table*}

\subsection{Mass outflow rate}
\label{sub:mass_outflow_rate} 
The mass outflow rate of the wind, $dM/dt=\mout$, is a crucial parameter and is the main means through which the overall flow energetics are assessed. The mass outflow rate for an arbitrary wind is given by $\mout = A(r)\rho(r)v(r)$, where $\rho(r)$ and $v(r)$ are the density and velocity profile of the wind and $A(r)$ is a factor which accounts for the geometry of the system. Assuming that the flow has constant terminal velocity $v(r)=\vout$ and that the absorbing gas has cosmic elemental abundances (i.e., $\sim75\%$ of its mass by hydrogen and $\sim25\%$ by helium), $\rho(r) \simeq 1.2 m_{\rm p} n(r)$ where $m_{\rm p}$ is the proton mass and $n(r)$ is the electron number density of the plasma. For a thin spherically symmetric isotropic wind $A(r)=\Omega b r^{2}$, where the product $\Omega b\leq 1$ is known as the \emph{global filling factor} and accounts for both the solid angle occupied by the flow ($\Omega$) and how much of the flow volume is filled by gas ($b$). Thus, $\mout \sim \Omega b r^{2} m_{\rm p} n(r) \vout$ where we have neglected the constant factor of order unity.

The value $b$ is extremely difficult to determine because it depends on the ionisation and clumpiness of the gas. At low--intermediate ionisation states the flow is likely to be clumpy/filamentary, while at high ionisation states it can be considered largely smooth and of low density because the vast majority of elements are stripped of electrons. In the clumpy case, the column density of the wind can be given by $\nh\sim b n(r) \delta r$, with $b<1$ implicitly allowing for inhomogeneities in the flow. Alternatively, at high ionisation states $b\simeq1$ and $r\to\rmax$. Substituting for $\rmin$ and $\rmax$ then leads to algebraic upper and lower limits on the mass outflow rate:
\begin{subequations}
\begin{align}
\moutmax & \sim \Omega m_{\rm p} \lion \xi^{-1} \vout,\label{eq:moutmax} \\
\moutmin & \sim 2 \Omega G \mbh m_{\rm p} \nh \vout^{-1}.\label{eq:moutmin}
\end{align}
\end{subequations}
The major remaining uncertainty is the opening angle of the system $\Omega$. Here, we adopt the average opening angle $\Omega=1.6\pi$ as inferred from the global detection fraction of Fe\,K winds reported in the literature ($f\simeq40\%$; \citealt{tombesi2010a,tombesi2014}, Paper~1). The assumption of a uniform wind geometry is clearly an oversimplification of real systems; in reality, numerous factors such as source luminosity, gas density, etc., will contribute to the shaping of an X-ray disk-wind and therefore the wind geometry is going to differ on an object-by-object basis. Even so, we note that the average opening angle of $\Omega=1.6\pi$ is comparable to the wind opening  measured from the P-Cygni-like Fe\,K profile in PDS\,456 ($\Omega$ is resolved to be $>2\pi$; see \citealt{nardini2015}). This suggests that it is a good approximation to the geometry of real disk-winds.

\subsection{Kinetic power}
\label{sub:kinetic_power}
Provided that the wind has already reached a steady terminal velocity by the point at which it is observed, the mechanical power imparted by expelling mass at a rate $\mout$ with velocity $\vout$ is simply equal to its kinetic energy: $\lk=\mout\vout^{2}/2$. We determine the range of likely kinetic power by substituting for $\mout=(\moutmax, \moutmin)$, leading to
\begin{subequations}
\begin{align}
\lkmax &\sim \Omega  m_{\rm p} \lion \xi^{-1} \vout^{3}\label{eq:kemax} \\
\lkmin &\sim G \Omega  m_{\rm p} \mbh \nh \vout,\label{eq:kemin}
\end{align}
\end{subequations}
Similarly, the rate at which the outflow transports momentum into the environment of the host galaxy is given by $dp/dt\equiv \dot p_{\rm out}=\mout\vout$. Substituting again for $\mout$ leads to a plausible range between
\begin{subequations}
\begin{align}
\poutmax &\sim \Omega m_{\rm p} \lion \xi^{-1} \vout^{2}\label{eq:poutmax} \\
\poutmin &\sim G \Omega m_{\rm p} \mbh \nh,\label{eq:poutmin}
\end{align}
\end{subequations}

\subsection{Other parameters}
\subsubsection{Black hole masses}
\label{sub:black_hole_masses}
Our method of estimating the inner radius of the outflow, $\rmin$, is proportional to the escape radius of the black hole, $\resc$, which in turn is proportional to the mass of the central black hole (BH), $\mbh$. By extension, this also means that the other lower limiting quantities in equations~(\ref{eq:rmin}), (\ref{eq:moutmin}), (\ref{eq:kemin}) and (\ref{eq:poutmin}) are also proportional to $\mbh$. Estimates for $\mbh$ gathered from the literature are collated in Table~\ref{table:measured_wind_properties}. We obtain most of the $\mbh$ values through the numerous reverberation mapping studies available in the literature (i.e., \citealt{peterson2004, bentz2009a, bentz2009b, edri2012}) which tend to offer relatively tight constraints on the mass of the central object; in the cases where an AGN had been subject to reverberation mapping multiple times, we adopt the most recent estimate on $\mbh$ only. There are several empirically determined mass estimates for the BHs in 3C\,111, ESO\,103-G035, NGC\,1365, NGC\,5506 and PDS\,456. For these AGN we report the mean value, with the associated errors taken as half of the range between the minimum and maximum values to account for the uncertainty in the individual estimates on $\mbh$. For CBS\,126, where we were unable to find any robust mass estimates in the literature, we infer $\mbh$ from the BLR line width and luminosity scaling relation of \citet{kaspi2000} using the appropriate spectral values listed in \citet{grupe2004}. For 3C\,390.3 we use the range of values inferred by \cite{dietrich2012}. In all other sources we take $\mbh$ directly from the reference listed in the table footnote. The sample encompasses almost 6 orders of magnitude in black hole mass, with $\log\mbh$ ranging from $\sim4.7$ in the dwarf-Seyfert galaxy NGC\,4395 (\citealt{edri2012}), all the way up to an estimated $\sim10$ in the high luminosity BAL quasar APM\,08279+5255 (\citealt{saez2009}). 

\subsubsection{Bolometric luminosity and Eddington ratio}
\label{ssub:bolometric_luminosity_eddr}
We estimate the bolometric luminosity for each AGN: $\lbol=\lcorr L_{\rm 2-10\,keV}$, where $L_{\rm 2-10\,keV}$ is the unattenuated source luminosity integrated between $2-10$\,keV and $\lcorr$ is the bolometric correction factor. Various studies have shown that the spread of $\lcorr$ amongst individual AGN is quite large (e.g., \citealt{elvis1994}), whilst the correction appropriate for a particular AGN can be a function of luminosity (\citealt{marconi2003,hopkins2007}), Eddington ratio (\citealt{vasudevan2007,lusso2010}), or both (\citealt{lusso2010}), such that the uniform application of a single bolometric correction factor may be inappropriate for a heterogeneously selected sample. We therefore primarily use the $\lcorr$ values listed in the works of \citet{vasudevan2007,vasudevan2009b} and \citet{vasudevan2010} which are empirically determined on the basis of the broad-band spectral energy distributions of each AGN. Several of the AGN in the sample are not listed in these works. For these sources, we searched the literature for an appropriate bolometric luminosity and used that instead (see Table~\ref{table:measured_wind_properties} caption). Only for 3C\,111 and CBS\,126 were we unable to locate either a bolometric correction factor or an definite and empirically measured bolometric luminosity; for these two sources we simply assume that $\lcorr\sim30$, which is similar to that found previously for other sources which harbour UFOs (e.g., \citealt{tombesi2012a}. Values for $L_{\rm 2-10\,keV}$ (hereafter denoted $L_{\rm X}$), $\lcorr$ and the resultant estimate of $\lbol$ are all listed in Table~\ref{table:normalised_table}. For completeness, we also compute the likely Eddington ratio of each AGN: $\eddratio\equiv\lbol/\ledd$, where $\ledd=4 \pi G m_{p} \mbh c \sigma_{T}^{-1}\simeq1.26\times10^{38}(\mbh/M_{\odot})$.

\section{Data Analysis} % (fold)
\label{sec:data_analysis}

\subsection{Preparing the data} % (fold)
\label{sub:preparing_the_data}
Several of the AGN in the have Fe\,K absorption detected in multiple epochs (see Paper~I), such as is the case for 3C\,111, Mrk\,766, NGC\,1365, NGC\,3227, NGC\,3783, NGC\,4051, or have more than one absorption trough which comprises a multi-velocity system (PDS\,456 and APM\,08279+5255). For these AGN considering the mean parameter value and accounting for intrinsic outflow variability between epochs by folding the uncertainties on each of the individual measurements into the error bar. In the majority of cases we find the individual measurements to be largely consistent within the errors, suggesting that the outflow is persistent across the different epochs. Only in 3C\,111, Mrk\,766 and NGC\,4051 do there appear to be significant differences at the 90\% level (see Table~\ref{table:measured_wind_properties}), and this is reflected in the broad error bars for these sources. For PDS\,456 and APM\,08279+5255, which both appear to have two Fe\,K absorption systems at different velocities, we adopt the mean velocity and again fold the range of possible velocities into the error bar. Table~\ref{table:measured_wind_properties} summarises the measured parameters of the winds detected in the \suzaku sample.

\subsection{Parameter constraints}
\label{sec:parameter_constraints}
Using the values in Table~\ref{table:measured_wind_properties} we first computed the wind parameters in standard units and then normalised them to the appropriate values for a given black hole mass; the resultant values are noted in Tables~\ref{table:main_results} and \ref{table:normalised_table}, respectively.We normalised the specific parameters as follows: (i) the distance $(\rmax,\rmin)$ to units of Schwarzchild radius $\rs=2G\mbh/c^{2}$, (ii) the kinetic luminosity $(\lkmax,\lkmin)$ to $\ledd$, (iii) the mass outflow rate $(\moutmax,\moutmin)$ to the Eddington accretion rate $\medd=\ledd/\eta c^{2}$ (assuming $\eta=0.06$ for the accretion efficiency), and (iv) the momentum rate to the Eddington momentum rate $\pedd=\ledd/c$. Algebraically, this yields a set of 8 normalised equations:
\begin{subequations}
\begin{align}
  \rmax/\rs	&= (2 G \mbh \xi \nh n_{\rm e})^{-1} n_{\rm H} \lion c^{2},\\
  \rmin/\rs	&= (c/\vout)^{2},\\
  \moutmax/\medd &= (4\pi G \xi \mbh n_{\rm e})^{-1} \Omega \sigma_{\rm T} n_{\rm H} \lion \vout \eta c,    \\
  \moutmin/\medd &= (2\pi\vout n_{\rm e})^{-1} \Omega \sigma_{\rm T} n_{\rm H}\nh \eta  c,   \\
  \lkmax/\ledd &= (4\pi G \mbh c \xi n_{\rm e})^{-1} \Omega \sigma_{\rm T} n_{\rm H}\lion \vout^{3},  \\
  \lkmin/\ledd &= (4\pi c n_{\rm e})^{-1} \Omega \sigma_{\rm T} n_{\rm H}\nh \vout,  \\
  \poutmax/\pedd &= (4\pi G \mbh \xi n_{\rm e})^{-1} \Omega \sigma_{\rm T} n_{\rm H} \lion \vout^{2}, \\
  \poutmin/\pedd &= (4\pi n_{\rm e})^{-1} \Omega \sigma_{\rm T} \nh.
\end{align}
\end{subequations}
Errors on the normalised parameters were determined by both propagating the uncertainties on the measured \xstar values through the various equations, and also taking into account the error on $\mbh$ where applicable. For CBS\,126, NGC\,1365, NGC\,3227, NGC\,3516 and NGC\,4395, which all have relatively slow wind velocities of the order $\vout<0.01$\,c (see Table~\ref{table:measured_wind_properties}), we found that $\rmin>\rmax$ (within the errors) which indicates that either: (i) the outflow may not have achieved the requisite escape speed for its observed location, in which case the wind material may not escape into the host galaxy unless it is subject to an additional acceleration mechanism once it has crossed the LOS, or (ii) that the wind steam line flows almost perpendicularly along the LOS such that we are only seeing the tangential component of $\vout$.  While the associated error bars in these slow sources may overlap (suggesting that the upper and lower limits may be formally consistent) we choose to only report the upper limiting quantities for these five AGN in Tables~\ref{table:main_results} and \ref{table:normalised_table} so that we do not introduce an artificially tight (and potentially misleading) constraint on the properties of the wind. In the remainder of our analysis we separate the outflow sample into two groups based on their velocity: those with $\vout\leq0.01$\,c ($3000$\,km\,s$^{-1}$) are hereafter classified as `slow', whilst those with $\vout>0.01$\,c are `fast'. Whilst the chosen velocity threshold is essentially arbitrary, this classification scheme provides a useful means to distinguish between those pristine `disc-winds' from those which may just be the higher-ionisation component of the inner-BLR or the more distant warm absorber, with which they probably share an overlap in velocity space and location (e.g., see \citealt{tombesi2012c}). 

\begin{table*}
	\begin{center}
	\begin{minipage}{19cm}
	\begin{threeparttable}
			\caption[Summary of normalised wind parameters]{Summary of normalised wind parameters.}
			\label{table:normalised_table}
	\footnotesize
		\begin{tabular}{l r c c c r r r r r r}% r}
		\toprule

		\multicolumn{1}{c}{\multirow{2}{*}{Source}}& 
		\multicolumn{1}{c}{\multirow{2}{*}{$\log L_{\rm X}^{\dag}|\lbol$} } & 
		\multicolumn{1}{c}{\multirow{2}{*}{$\lcorr$} } &
		\multicolumn{1}{c}{$\log \rmin$}  & 
		\multicolumn{1}{c}{$\log \rmax$}  & 
		\multicolumn{1}{c}{$\log \moutmin$}  & 
		\multicolumn{1}{c}{$\log \moutmax$} & 
		\multicolumn{1}{c}{$\log \lkmin$}  & 
		\multicolumn{1}{c}{$\log \lkmax$}\\[0.5ex]
 
		&
		&
		&
		\multicolumn{1}{c}{$(\rs)$} & 
		\multicolumn{1}{c}{$(\rs)$}  & 
		\multicolumn{1}{c}{$(\medd)$}  & 
		\multicolumn{1}{c}{$(\medd)$}  & 
		\multicolumn{1}{c}{$(\ledd)$} & 
		\multicolumn{1}{c}{$(\ledd)$}\\
	
		\midrule
	
		3C\,111  	& $44.35\pm0.18~|~45.7$ & $20^{\,6}$ 
				 	& $2.3\pm0.1$ & $3.7\pm0.6$ & $-1.3\pm0.1$ & $ 0.1\pm0.6$ & $-2.9\pm0.1$ & $-1.5\pm0.6$ \\[0.5ex]
	
		3C\,390.3  	& $44.42\pm0.01~|~45.5$ & $13.3^{\,3}$ & $1.7\pm0.1$ 
					& $<2.8$ & $>-1.1$ & $<-0.0$ & $>-2.1$ & $<-2.8$ \\[0.5ex]
	
		4C\,+74.26 	& $44.95\pm0.01~|~46.3$ & \na$^{\,?}$ & $1.5\pm0.1$ 
					& $<5.0$ & $>-2.8$ & $0.2\pm0.5$ & $>-3.8$ & $-0.6\pm0.5$ \\[0.5ex]
	
		APM\,08279  & $46.73\pm0.03~|~47.4$ & \na$^{\,1}$ &  $1.1\pm0.1$ 
						& $4.4\pm0.3$ & $-1.9\pm0.1$ & $ 1.4\pm0.3$ & $-2.3\pm0.1$ & $ 1.0\pm0.3$ \\[0.5ex]
	
		CBS\,126  		& $43.82\pm0.02~|~45.1$ & $20^{\,?}$ & \na & $<2.6$ & \na 
						& $-1.1^{+0.2}_{-0.3}$ & \na & $-4.3^{+0.2}_{-0.3}$ \\[0.5ex]
	
		ESO\,103-G035  	& $43.47\pm0.01~|~44.6$ & $12.6^{\,4}$ & $2.5^{+0.5}_{-0.3}$ 
						& $<6.0$ & $>2.6$ & $0.0\pm1.1$ & $>-4.4$ & $-1.8\pm1.1$ \\[0.5ex]
	
		MCG\,-6-30-15  	& $42.77\pm0.01~|~44.0$ & $16.1^{\,5}$ & $4.3^{+0.3}_{-0.2}$ 
						& $5.23\pm0.1$ & $-1.1^{+0.3}_{-0.2}$ & $-0.2\pm0.1$ & $-4.7^{+0.3}_{-0.2}$ & $-3.8\pm0.1$ \\[0.5ex]
	
		MR\,2251-178  	& $44.61\pm0.01~|~46.0$ & $22.0^{\,3}$ &  $1.7\pm0.1$ 
						& $5.2\pm0.1$ & $-2.53\pm0.1$ & $1.0^{+0.5}_{-0.8}$ & $-3.55\pm0.1$ & $ 0.1^{+0.5}_{-0.8}$ \\[0.5ex]
	
		Mrk\,279  		& $42.78\pm0.01~|~44.0$ & $10.1^{\,5}$ & $1.3\pm0.1$ 
						& $2.6^{+0.6}_{-0.4}$ & $-1.36\pm0.1$ & $-0.1^{+0.6}_{-0.4}$ & $-1.98\pm0.1$ & $-0.7^{+0.6}_{-0.4}$ \\[0.5ex]
	
		Mrk\,766 		& $42.73\pm0.05~|~44.1$ & $48.9^{\,5}$ & $2.8\pm0.1$ 
						& $5.0\pm0.3$ & $-1.3\pm0.1$ & $ 0.9\pm0.3$ & $-3.4\pm0.1$ & $-1.2\pm0.3$ \\[0.5ex]

		NGC\,1365 		& $42.32\pm0.02~|~44.1$ & $101.6^{\,5}$ & \na 
						& $2.1\pm0.1$ & \na & $-1.7\pm0.1$ & \na & $-5.2\pm0.1$ \\[0.5ex]
	
		NGC\,3227 		& $42.13\pm0.06~|~43.3$ & $15.3^{\,5}$ & \na & $2.9\pm0.3$ 
						& $-0.4^{+1.4}_{-0.5}$ & $-2.2\pm0.3$ & \na & $-6.1\pm0.3$ \\[0.5ex]
	
		NGC\,3516  		& $43.07\pm0.01~|~44.1$ & $14.0^{\,5}$ & \na 
						& $4.1\pm0.3$ & \na & $-1.13\pm0.3$ & \na & $-5.23\pm0.3$ \\[0.5ex]
	 
		NGC\,3783  		& $43.14\pm0.05~|~44.5$ & $23.7^{\,5}$ & $>4.3$ 
						& $5.4\pm0.3$ & $>-1.5$ & $<-0.1$ & $<-5.1$ & $<-4.3$ \\[0.5ex]
	 
		NGC\,4051  		& $41.65\pm0.04~|~43.2$ & $41.1^{\,4}$ &  $3.5\pm0.1$ 
						& $3.9\pm0.2$ & $-0.9\pm0.1$ & $-0.4\pm0.2$ & $-3.7\pm0.1$ & $-3.2\pm0.2$ \\[0.5ex]
	 
		NGC\,4151  		& $42.34\pm0.02~|~43.6$ & $18.3^{\,4}$ &  $2.5^{+0.5}_{-0.3}$ 
						& $<5.6$ & $>-2.7$ & $0.0\pm0.6$ & $>-4.5$ & $-1.8\pm0.6$ \\[0.5ex]

		NGC\,4395 		& $40.38\pm0.01~|~41.7$ & $22.5^{\,3}$ & \na 
						& $3.8\pm0.4$ & \na & $<-1.4$ & \na & $<-7.5$ \\[0.5ex]
	
		NGC\,5506 		& $43.19\pm0.01~|~44.4$ & $16.8^{\,5}$ &  $1.2\pm0.1$ 
						& $2.6\pm0.5$ & $-1.6\pm0.1$ & $-0.2\pm0.5$ & $-2.1\pm0.1$ & $-0.7\pm0.5$ \\[0.5ex]
	
		PDS\,456  & $44.62\pm0.01~|~47.0$ & \na$^{\,2}$ & $1.1\pm0.1$ 
						& $3.3\pm0.2$ & $-1.8\pm0.1$ & $ 0.4\pm0.2$ & $-2.2\pm0.1$ & $0.0\pm0.2$ \\[0.5ex]
	
		SW\,J2127 & $43.15\pm0.01~|~44.5$ & \na$^{\,?}$ & $1.3\pm0.1$ 
						& $4.1\pm0.5$ & $-2.0\pm0.1$ & $ 0.8\pm0.5$ & $-2.6\pm0.1$ & $ 0.3\pm0.5$ \\
		\bottomrule
		\end{tabular}
			\begin{tablenotes}
			\item[]\textsc{Notes:}-- $^{1}$The bolometric luminosity of APM\,08279+5255 is taken as $\lbol=7\times10^{15}\mu_{L}^{-1}\,L_{\odot}$ (\citealt{lewis1998, riechers2009, saez2011}), assuming a conservative magnification factor of $\mu_{L}=100$ (\citealt{egami2000}); $^{2}\lbol$ for PDS\,456 taken from \citet{reeves2000}; $^{3,4,5}\kappa_{\rm bol}$ correction factors taken from \citet{vasudevan2007, vasudevan2009a} or \citet{vasudevan2010}, respectively; $^{6}\kappa_{\rm bol}$ estimated from the $\Gamma-\kappa_{\rm bol}$ correlation of \citet{zhou2010} using the best-fit $\Gamma$ values that we found in Paper~I. $^{\dag}$$L_{\rm X}$ here corresponds to the unattenuated source luminosity between $2-10$\,keV.
			\end{tablenotes}
	\end{threeparttable}
	\end{minipage}
	\end{center}
\end{table*}

\begin{figure*}
\begin{minipage}{\textwidth}
	\begin{center}
		\includegraphics[width=0.95\textwidth]{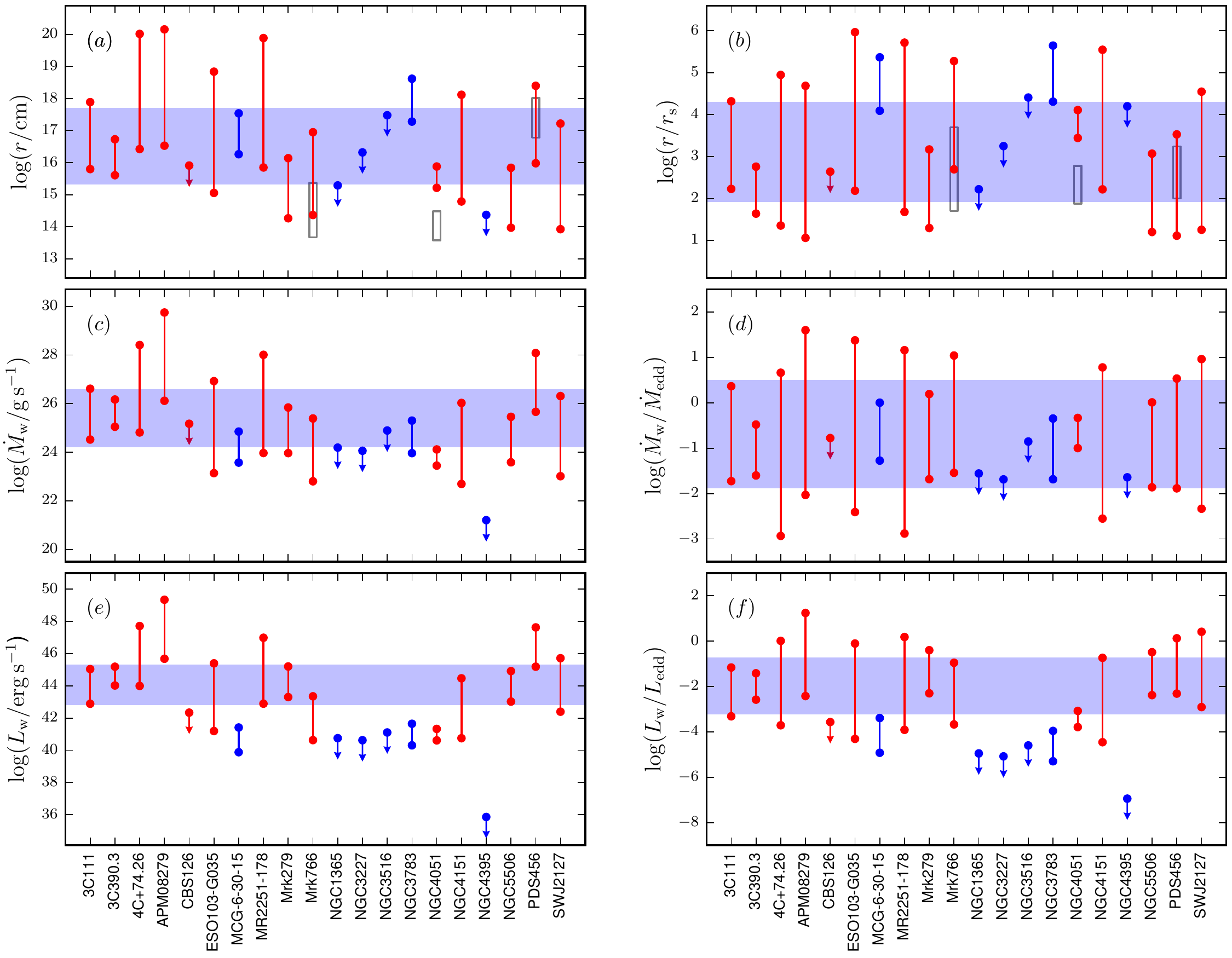}

	\end{center}
	\caption{Constraint diagram showing the parameter ranges occupied by the \suzaku-detected Fe\,K outflows. The left hand side panels shows the raw limiting values (taken from Table~\ref{table:main_results}) and the right hand side shows the normalised ones (from Table~\ref{table:normalised_table}). The top, middle and bottom rows are the constraints on distance, mass outflow rate and kinetic power. In all panels, the red and blue lines show sources with $\vout>3000$\,km\,s$^{-1}$ and $\vout<3000$\,km\,s$^{-1}$, respectively. For Mrk\,766, NGC\,4051 and PDS\,456, the grey boxes show the range in location measured by \citet{risaliti2011,miller2007,turner2007,miller2010b} and \citet{gofford2014} on the basis of variability or X-ray reverberation studies, respectively (see text for further details). The blue/shaded areas show the `mean range', spanning the mean of all upper and all lower limiting values of the fast winds only.}
	\label{fig:constraint_diagrams}
\end{minipage}
\end{figure*}

In Figure~\ref{fig:constraint_diagrams} we show constraint diagrams for the various wind parameters\footnote{Note that we do not show a constraint diagram for $\pout$ because it is simply proportional to $\lk/\vout$ and therefore follows a similar overall distribution to the one for $\lk$}. From Figure~\ref{fig:constraint_diagrams}a--b it is clear that the absorbing material is distributed across several orders of magnitude in distance from their black hole, with $r$ spanning $\sim10^{14-20}$\,cm or $\sim10^{1-6}$\,$\rs$. By considering the mean range\footnote{Here, `mean range' refers to the range spanning between the means for the upper and lower limit values, respectively, i.e., it corresponds to the range between between $\langle\rmax\rangle \rightarrow \langle \rmin \rangle$ for the fast outflows (see caption for more details)} we see that the fast outflows appear to be located between $\langle r \rangle \sim10^{15-18}$\,cm ($\sim10^{2-4}$\,$\rs$) distance range. This corresponds to $\sim0.0003-0.3$\,pc and implies that whilst there may be some limited overlap between the fast Fe\,K outflows and the traditional soft X-ray warm absorber (which is usually inferred to be on parsec-scale distances, e.g., \citealt{blustin2005, kaastra2012, crenshaw&kraemer&george2003} and references therein), the former are generally located much closer to their central black hole. This is in agreement with other results in the literature which find the highly-ionised Fe\,K outflows most likely originate in a wind from the inner regions of the accretion disc (e.g., \citealt{proga2004,schurch2009,sim2008,king2010}), and is also consistent with what \citet{tombesi2012a} found on the basis of the \xmm outflow sample.

Following our estimates on $r$, Figure~\ref{fig:constraint_diagrams}c shows an analogous plot for $\mout$. The constraints on $\mout$ also vary significantly, spanning almost 8 orders of magnitude, and ranging from $<10^{22}$\,g\,s$^{-1}$ in the dwarf Seyfert galaxy NGC\,4395 all the way up to $\sim10^{30}$\,g\,s$^{-1}$ in the massive quasar APM\,08279+5255. The other winds have $\mout$ estimates distributed between these two extremes, with the mean range for the fast outflows falling between $\langle \mout \rangle \sim 10^{24-26}$\,g\,s$^{-1}$ ($\sim0.01-1\,M_{\odot}$\,yr). Interestingly, and while there remains considerable uncertainty in each case, $\langle \moutmax \rangle$ is of the order of the Eddington rate, i.e., $\langle \moutmax \rangle\approx\medd$, while even $\langle \moutmin \rangle$ is still $\sim1\%$ of $\medd$. This immediately suggests that the winds are transporting a substantial amount of material into their host galaxies. The ejected mass has a large kinetic power: Figure~\ref{fig:constraint_diagrams}e shows that while the range of $\lk$ again spans several orders of magnitude, the overall mean for the fast flows is constrained to between $\langle \lk \rangle \sim 10^{43}$ and $10^{45}$\,erg\,s$^{-1}$. The normalised values show that, on average, $\langle \lk \rangle$ is $<\ledd$, with a mean range of $\langle \lk \rangle \sim (0.1-10\%)\ledd$. The wind mass outflow rate and the associated kinetic power are therefore significant fractions of the Eddington limited values. In \S\ref{subsec:energetic_significance} we extend these results to assess whether the wind is energetically significant in terms of feedback.
 
As mentioned previously, our estimates on $r$ are subject to large uncertainties. It can therefore be useful to compare a few of our inferred values with some more robust measurements available in the literature. We consider here the exemplar cases of PDS\,456 and Mrk\,766. Currently, PDS\,456 is the only AGN where the location of the Fe\,K-shell wind has been robustly constrained from discrete line variability. On the basis of a recent \suzaku campaign \citet{gofford2014} constrained the Fe\,K absorber in PDS\,456 to $\sim100-1700$\,$\rs$. This is entirely consistent with what we infer using equations~(\ref{eq:rmax}--\ref{eq:rmin}). Similar is also true for Mrk\,766: \citet{miller2007} and \citet{turner2007} estimates the absorbing gas to be located on distances of the order $30-50$\,$\rs$, while \citet{risaliti2011} later showed that both the broad-band spectral variability and the Fe\,K absorption could also be attributed to a stratified, inhomogenous and clumpy absorber which occults the AGN at a distance of $\sim500-5000$\,$\rs$. These estimates are again consistent with what we find here (see Figure~\ref{fig:constraint_diagrams}a-b), suggesting that the two studies are perhaps probing the same layer of gas. Elsewhere in the literature, \citet{miller2010b} used a detailed spectral timing analysis to de-convolve the X-ray spectrum of NGC\,4051 into its constituent absorption- and reflection-dominated components, with the hard X-ray reverberation signal suggesting that the reflecting gas is $\sim100-600$\,$\rs$ from the black hole (at the 90\% level, see \citealt{miller2010b}). This is around one order of magnitude closer than where we estimate the Fe\,K absorbing gas to be in this source. One possibility is that the reflecting layer detected by \citealt{miller2010b} is physically distinct to the Fe\,K absorber, possibility associated with the Compton-thick occluding clouds posited by \citet[2014, in prep]{tatum2013}.

\begin{figure}
	\begin{center}
		\includegraphics[width=0.45\textwidth]{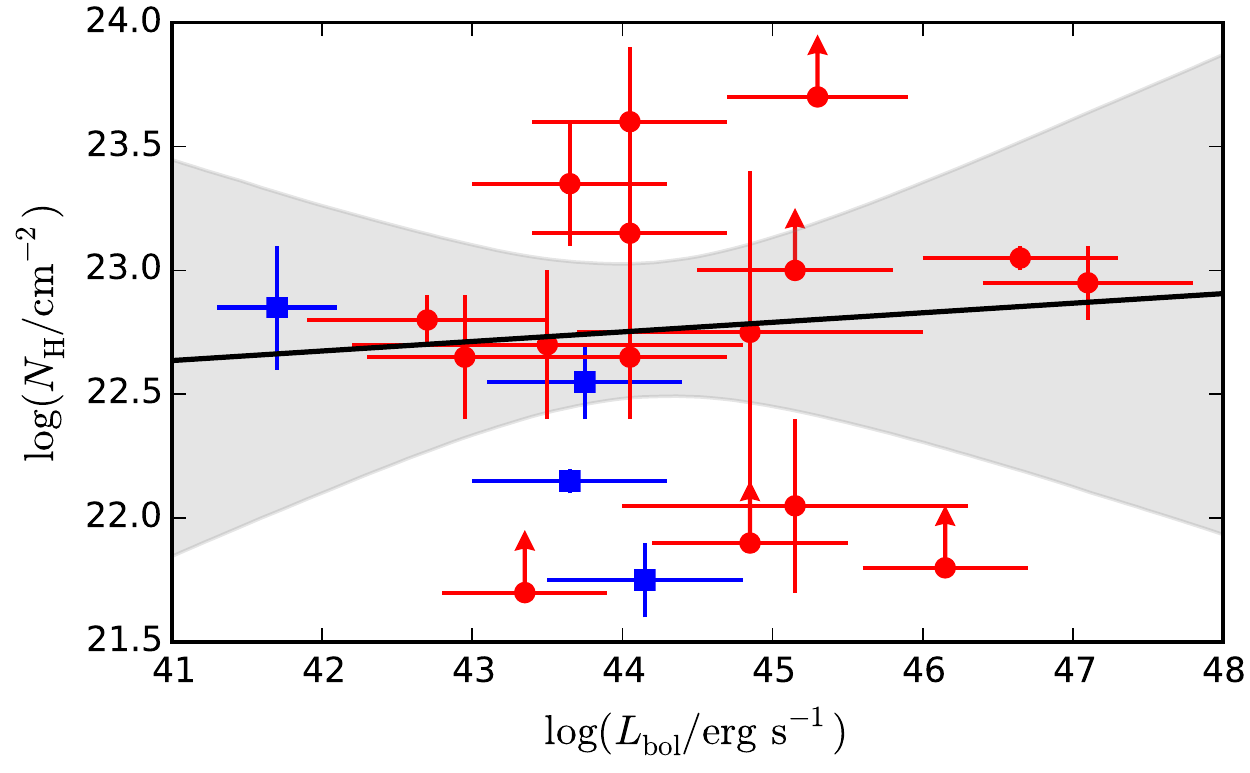}
		\includegraphics[width=0.45\textwidth]{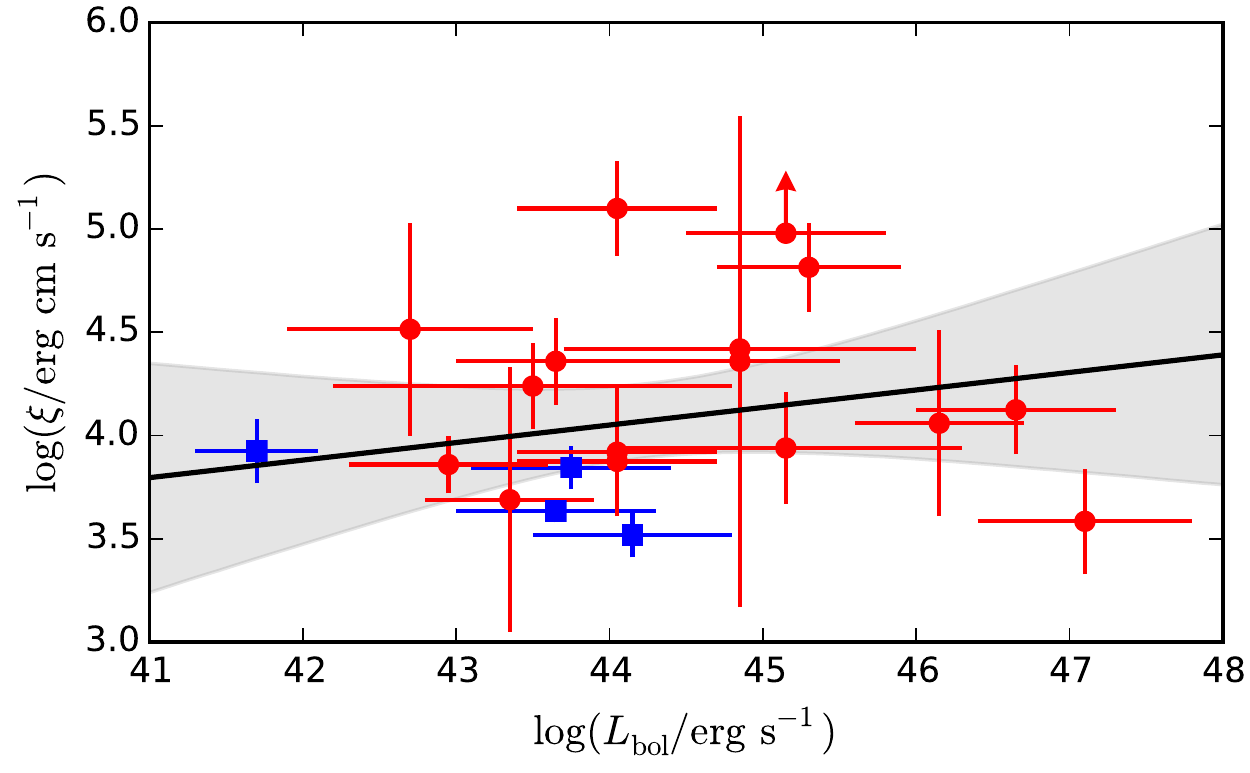}
		\includegraphics[width=0.45\textwidth]{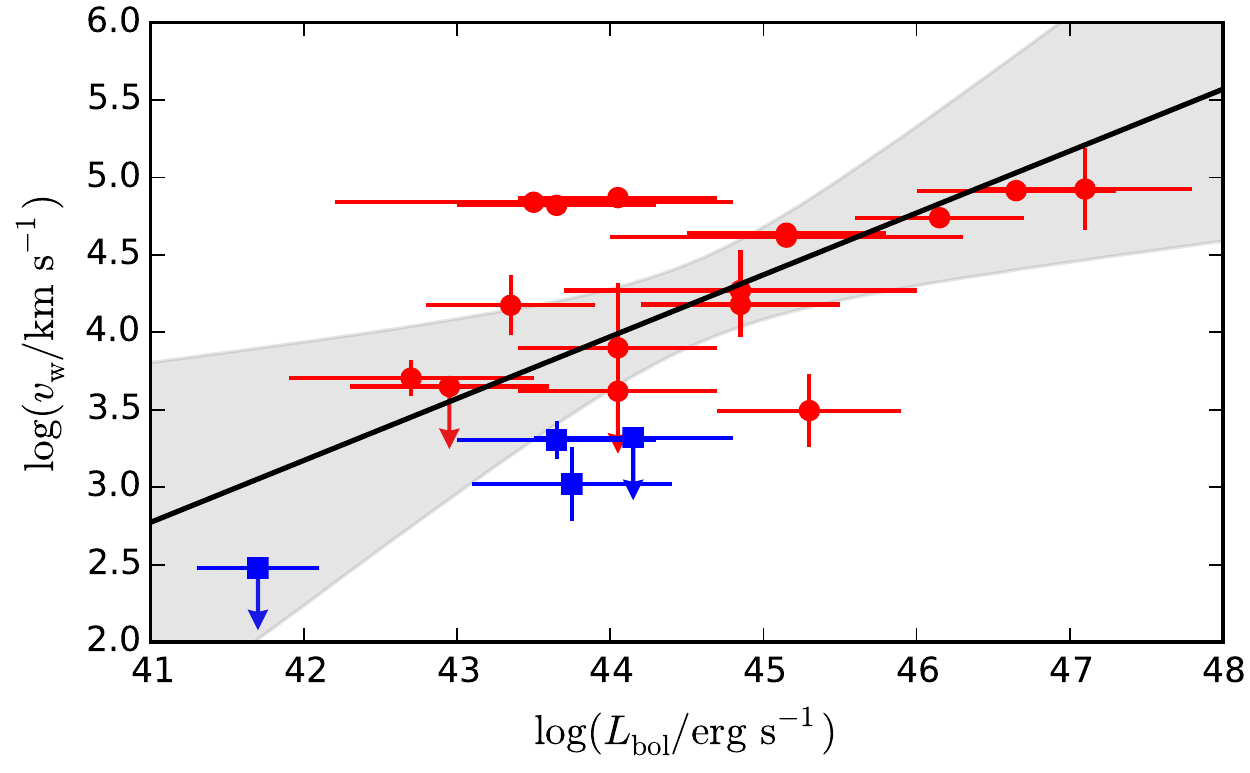}
	\end{center} 
	\caption{Scatter plots showing $\log\lbol$ versus: (a) $\lognh$, (b) $\logxi$ and (c) $\log(\vout/{\rm km\,s}^{-1})$, as noted in Table~\ref{table:main_results}. The red and blue data-points correspond to AGN with $\vout>3000$\,km\,s$^{-1}$ and $\vout<3000$\,km\,s$^{-1}$, respectively. The solid black line corresponds to the `best-fit' linear regression to the fast winds, as estimated from 100,000 MCMC realisations of the dataset using the {\tt linmix\_err} Bayesian regression algorithm (\citealt{kelly2007}), while the grey shaded area denotes the region which contains 90\% of the posterior probability distribution. The Parameters of the fits are reported in Table~\ref{tab:linear_regression}. }
	\label{fig:lbol_absorber}
\end{figure}

\subsection{Correlations with bolometric luminosity}
\label{sec:correlation_analyses}
Supplementing the constraint diagrams we also carried out a correlation analysis. This analysis bears notable similarities to the one recently conducted by \citet{tombesi2012c} for the \xmm outflow sample. There, the authors combined the parameters of the hard-band absorber, as measured in their Fe\,K absorption survey with \xmm, with those for the soft X-ray warm absorber collated from the literature to determine how the AGN wind parameters vary globally in relation to their distance from the black hole (\citealt{tombesi2012c}). Here, noting that \suzaku does not have the soft X-ray energy resolution necessary to constrain the velocity of the soft X-ray absorbing gas, we consider only the Fe\,K absorbers and do not consider the warm absorber. Whilst this leads to a relatively small dynamic range in terms of their distance from the black hole, the heterogeneous nature of the \suzaku sample encompasses a significantly broader range of AGN in terms of their bolometric luminosity ($\lbol$). The \suzaku sample therefore offers a useful opportunity to probe how the wind parameters may vary in comparison to the luminosity of their host AGN.

\begin{table}
 \begin{center}
 %\begin{minipage}{10cm}
  \caption{Summary of linear regression parameters: $\log(y)=\alpha+\beta\log(x)$}
  	\label{tab:linear_regression}
  \begin{threeparttable}
  \begin{tabular}{llrrrr}
  \toprule
  \multicolumn{2}{c}{Relation} & 
  \multicolumn{1}{c}{$\alpha$} & 
  \multicolumn{1}{c}{$\beta$} &
  \multicolumn{1}{c}{$\sigma$} &
  \multicolumn{1}{c}{$R_{p}$} \\ 
  \cmidrule(r{.75em}l){1-2}
	 
  \multicolumn{1}{c}{$\log (x)$} & 
  \multicolumn{1}{c}{$\log (y)$} &
  \multicolumn{1}{c}{(1)}&
  \multicolumn{1}{c}{(2)}&
  \multicolumn{1}{c}{(3)}&
  \multicolumn{1}{c}{(4)}\\
  \midrule
  $\lbol$ 
   & $\nh^{a}$ & $21.1^{+10.7}_{-10.5}$ & $0.0^{+0.3}_{-0.2}$ & $0.6^{+0.3}_{-0.2}$ & $0.11^{+0.59}_{-0.65}$\\[0.5ex]

   & $\xi^{a}$ & $0.3^{+7.1}_{-7.2}$ & $0.1^{+0.2}_{-0.2}$ & $0.4^{+0.6}_{-0.1}$ & $0.32^{+0.49}_{-0.60}$\\[0.5ex]

   & $\vout$ & $-13.6^{+12.4}_{-15.4}$ & $0.4^{+0.3}_{-0.2}$ & $0.6^{+0.3}_{-0.2}$ & $0.69^{+0.26}_{-0.47}$\\

  \cmidrule{2-6}
   & $r$ & $-8.4^{+38.3}_{-35.9}$ & $0.6^{+0.8}_{-0.7}$ & $0.8^{+0.8}_{-0.6}$ & $0.71^{+0.28}_{-0.93}$\\[0.5ex]

   & $\mout$ & $-13.0^{+28.1}_{-35.4}$ & $0.9^{+0.8}_{-0.6}$ & $0.6^{+0.7}_{-0.4}$ & $>0.34$\\[0.5ex]

   & $\lk$ & $-23.5^{+23.6}_{-44.7}$ & $1.5^{+1.0}_{-0.8}$ & $1.0^{+1.2}_{-0.7}$ & $>0.54$ \\[0.5ex]

   & $\pout$ & $-18.1^{+28.8}_{-36.9}$ & $1.2^{+0.8}_{-0.7}$ & $0.7^{+0.9}_{-0.5}$ & $>0.55$\\		
  \cmidrule{2-6}
   & $r/\rs$ & $24.1^{+84.0}_{-34.3}$ & $-0.5^{+0.8}_{-0.8}$ & $0.8^{+0.9}_{-0.6}$ & $-0.60^{+0.99}_{-0.38}$\\[0.5ex]

   & $\mout/\medd$ & $0.4^{+29.5}_{-29.6}$ & $0.0^{+0.7}_{-0.7}$ & $0.5^{+0.7}_{-0.1}$ & $0.04^{+0.90}_{-0.99}$\\[0.5ex]

   & $\lk/\ledd$ & $-27.0^{+34.0}_{-35.2}$ & $0.6^{+0.8}_{-0.8}$ & $0.8^{+0.9}_{-0.6}$ & $0.75^{+0.24}_{-1.02}$\\[0.5ex]

   & $\pout/\pedd$ & $-12.4^{+30.5}_{-30.9}$ & $0.3^{+0.7}_{-0.7}$ & $0.6^{+0.8}_{-0.4}$ & $0.56^{+0.43}_{-1.32}$\\

  \midrule
  $\pbol$ & $\pout$ & $-5.4^{+21.6}_{-27.9}$ & $1.2^{+0.8}_{-0.7}$ & $0.7^{+0.9}_{-0.5}$ & $>0.55$\\

  \bottomrule
  \end{tabular}
  	\begin{tablenotes}
  	  	\item[]\textsc{Notes:} All regressions were performed using the {\tt linmix\_err} Bayesian regression routine (\citealt{kelly2007}) which takes into account both measurement and upper limits in the $y$-variable. The noted best-fit values correspond to the median value of the Posterior probability distribution, as simulated from 100,000 MCMC realisations of the data, with errors taken as the range of values which encompass 90\% of the Posterior probability. See \citet{kelly2007} for more details. Nominal values correspond to a fit to the entire sample, while those in brackets are for an analogous fit with NGC\,4395 removed. $^{a}$Regression computed with lower limits excluded.\\
  		\item[]\textsc{Columns:} (1) slope of the linear regression; (2) normalisation/intercept of the best-fit regression line; (3) standard deviation of the intrinsic scatter in the data; (4) best-fit linear correlation coefficient. $\pm1$ denote perfect positive/negative correlations, respectively.
  	\end{tablenotes}
  \end{threeparttable}
  %\end{minipage}
  \end{center}
\end{table}

We compute linear regressions of the form $\log(y)=\alpha+\beta\log(x)$, where $\alpha$ and $\beta$ are the intercept and slope of the straight line fit, respectively, using the {\tt linmix\_err} Bayesian regression algorithm\footnote{available in IDL from: } (\citealt{kelly2007}). This routine employs Markov Chain Monte Carlo (MCMC) techniques to self-consistently account for measurement errors and intrinsic scatter in the data, whilst also allowing for limited censorship (upper-limits only) in the independent variable. We computed regression parameters from 100,000 MCMC realisations of the data by the {\tt linmix\_err} routine. This yields posterior distributions for the intercept $\alpha$, slope $\beta$, standard deviation (scatter) of the data $\sigma$, and the Pearson linear correlation coefficient $R_{p}$. We adopt the median value of the posterior distribution as our `best-fit' to the data, and estimate the parameter errors as the range of simulated values which encompass 90\% of the posterior distribution about the median. 

We note the slow winds have $\vout\leq3000$\,km\,s$^{-1}$ which is similar to that measured in the traditional soft X-ray warm absorber. It is therefore unclear whether these winds are associated with a {\it bona fide} disk-wind or alternatively with another layer of gas which is more distant from the black hole, e.g., with a higher ionisation component of the warm absorber. For this reason, we fit all subsequent regression analyses to the fast winds only, $\vout>3000$\,km\,s$^{-1}$. Owing to their high velocities, these faster winds are kinematically distinct from the soft X-ray absorbing gas and thus may more obviously represent the signature of a pristine disk-winds. This limits our analysis to the winds which are capable of escaping the gravitational potential of the central black hole (based on our LOS), and are therefore more likely to have an effect on the host galaxy in terms of feedback. 

In Figure~\ref{fig:lbol_absorber} we show how $\lognh$, $\logxi$ and $\log(\vout/{\rm km\,s}^{-1})$, as measured in Paper~I and summarised in Table~\ref{table:main_results}, vary with $\lbol$. 
From Figures~\ref{fig:lbol_absorber}a and \ref{fig:lbol_absorber}b we find no discernible relationships between either $\nh$ or $\xi$ and $\lbol$ ($\beta=0$ at 90\% confidence): the $\lognh$ values cover a wide range for a given value of $\lbol$, while $\logxi$ appears to cluster at around $\sim4$, mirroring the median of the parameter distributions that we found in Paper~I. This also appears to be true for both the fast and the slow wind sub-samples. Conversely, there is a  correlation present between $\vout$ and $\lbol$, with the faster winds tending to be being observed in more luminous AGN. The fast winds have a constrained slope of $\beta=0.4^{+0.3}_{-0.2}$, with $\vout\propto\lbol^{\beta}$ and $\beta\approx0.5$. In contrast, the slower systems, i.e., those with $\vout<0.01$\,c, appear to be isolated to the lower left quarter of the plot which suggests that low velocity winds are preferentially located in lower luminosity systems. This is consistent with the overall picture that wind velocity is correlated with $\lbol$; fitting a correlation to the entire sample yields $\beta_{\rm all}=0.5^{+0.4}_{-0.2}$. This is formally consistent within the errors but does suggest that the inclusion of the slow winds may skew the relationship to a slightly steeper slope.  

Figure~\ref{fig:lbol_raw} then shows how the raw wind parameters vary with $\lbol$. Here, and in subsequent figures, each data point corresponds to the mid-point in the range for each parameter, calculated as, e.g.,:
\begin{equation}
 	 r_{\rm mid} = \dfrac{(\rmax+{\tt perr})+(\rmin-{\tt nerr})}{2},
 	\label{eqn:mean_value}
\end{equation}
where {\tt nerr} and {\tt perr} are the negative and positive error, respectively. The associated error bars then denote the range between the maximum and minimum values, including the uncertainties arising from the input variables. In equation~\ref{eqn:mean_value} we are essentially only considering the upper- and lower-limits for {\it all} derived values in the sample, such that in the cases where the minimum and maximum values are themselves unconstrained (i.e., only and upper- or lower-limit is available with no associated error) we are able to use the censored value when calculating the mid-point. This ensures that all of the available data is included in our analysis.

To first order, we find that all of the raw wind parameters are plausibly correlated with $\lbol$ to some degree. Whilst the slope for $\lbol-r$ is tentative, and marginally consistent with $\beta=0$ at the 90\% level (Figure~\ref{fig:lbol_raw}, the overall distribution is skewed towards a positive relationship among the data which suggests that the winds are observed at larger radii in brighter AGN, as would be expected given the larger typical size-scales in these systems. Figure~\ref{fig:lbol_raw}) then shows that there is a positive relationship between $\lbol-\mout$ ($\beta=0.9^{+0.8}_{-0.6}$), such that the winds in more luminous AGN are correspondingly more massive. In both of these cases the slow winds appear to follow the same overall trend as the fast ones, despite not formally being included in the regression computation. Similar is also true for $\lbol-\lk$ and $\lbol-\pout$, which share strong positive slopes of $\beta=1.5^{+1.0}_{-0.8}$ ($R_{\rm p}\geq0.54$) and $\beta=1.2^{+0.8}_{-0.7}$ ($R_{p}\geq0.55$), respectively. The observed slopes here likely stem from their mutual dependence on $\vout$: $\lk\propto\vout^{3}$ and $\pout\propto\vout^{2}$. Indeed, the $\lk\propto\lbol^{\sim1.5}$ and $\pout\propto\lbol^{\sim1.0}$ relationships implied here are formally consistent with the $\vout\propto\lbol^{\sim0.5}$ found earlier in Figure~\ref{fig:lbol_absorber}c. This ultimately suggests that the wind velocity may be the driving factor in the observed relationships, with the uncertainties on the other wind parameters only adding to the underlying scatter of the data. In conclusion, these results indicate that disk-winds in higher luminosity sources are (tentatively) located farther away from their central black hole in terms of absolute distance (as expected for larger $\mbh$), they contain more material, and they are accordingly more energetic. The driving factor behind these relationships appears to be the wind velocity, which is faster in more luminous objects.

\begin{figure*} 
	\begin{center}
		\includegraphics[width=0.45\textwidth]{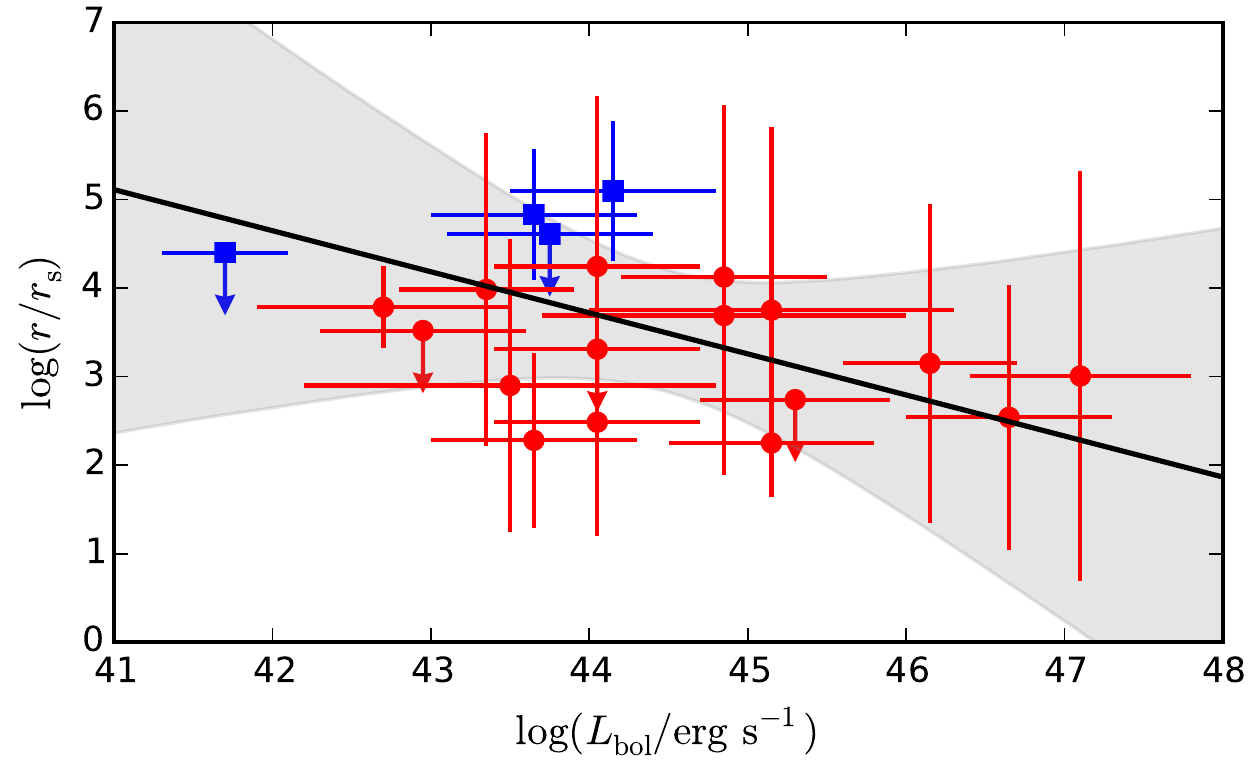}
		\includegraphics[width=0.45\textwidth]{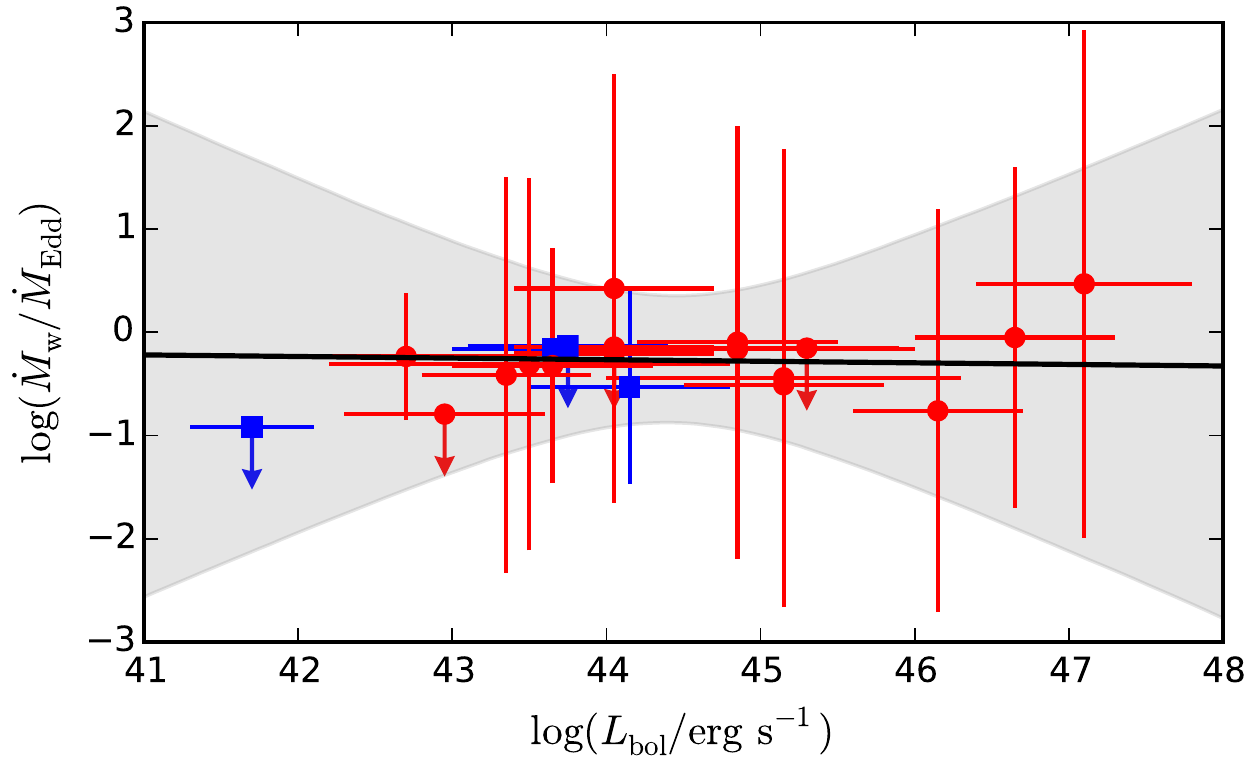}
		\includegraphics[width=0.45\textwidth]{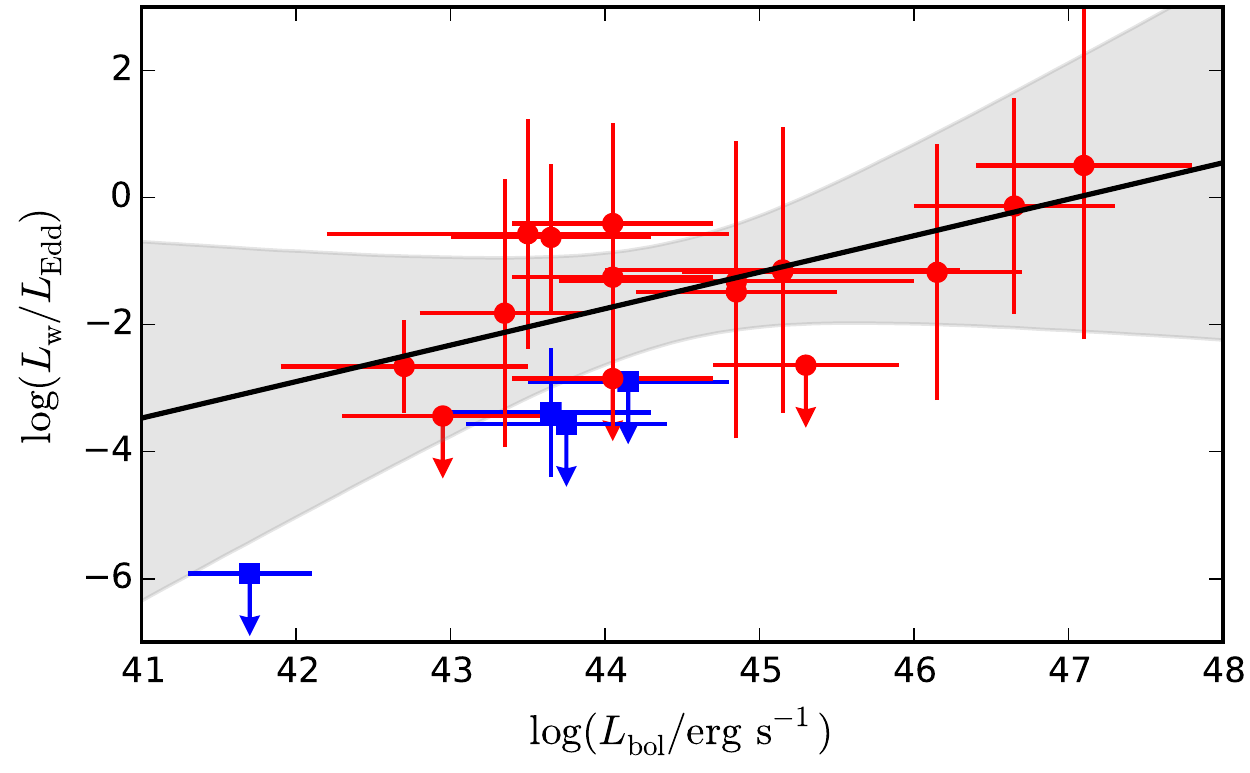}
		\includegraphics[width=0.45\textwidth]{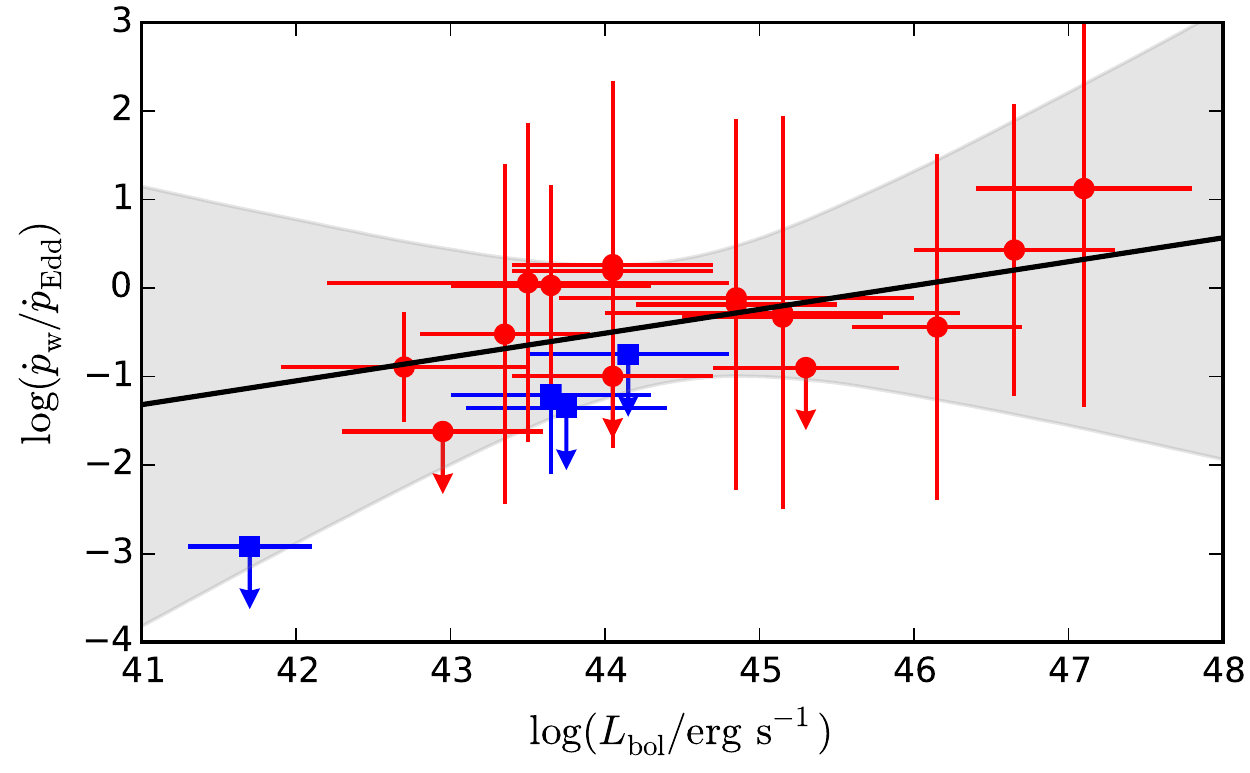}
	\end{center}
	\caption{More scatter plots, this time showing $\log\lbol$ against the normalised wind parameters: (a) $\log(r/\rs)$, (b) $\log(\mout/\medd)$, (c) $\log(\lk/\ledd)$ and (d) $\log(\pout/\pedd)$. Data points and plot components have the same meaning as in previous figures.}
	\label{fig:lbol_norm}
\end{figure*}

Similar scatter plots for the normalised wind parameters are shown in Figure~\ref{fig:lbol_norm}. In contrast to before, this time we are unable to discern any linear correlations in the normalised $\lbol-(r/\rs)$ and $\lbol-(\mout/\medd)$ plots. In fact, and whilst there is a weak negative correlation suggested in the case of the former, both of the relationships are formally consistent with $\beta=R_{p}=0$ such that the $r/\rs$ and $\mout/\medd$ ratios are largely independent of $\lbol$. The fast winds are therefore observed at similar relative distances from the SMBH regardless of the source luminosity, and their mass outflow rate is a similar fraction of $\medd$. The slow winds also appear to follow consistent relationships. Both $\lk/\ledd$ (Figure~\ref{fig:lbol_norm}d) and $\pout/\pedd$ (Figure~\ref{fig:lbol_norm}e) are also formally consistent with a slope of zero, although weak positive correlations are suggested. In both Figure~\ref{fig:lbol_norm}c and \ref{fig:lbol_norm}d, the slow winds appear to lie below the computed regression line for the fast winds. Their location in these plots appears to mirror that seen in Figure~\ref{fig:lbol_norm} for the $\lbol-\vout$ correlation, which is again consistent with the wind velocity being the parameter behind the relationships.   

For completeness, we also tried to determine whether the parameters of the wind scale with the Eddington ratio of the AGN. Unfortunately, however, the relatively tight clustering in the data at $\eddratio=0.1$, compounded by the lack of Fe\,K wind parameters in low Eddington ratio sources (i.e., $\eddratio\lesssim10^{-2}$) sources, meant that we were unable to discern any plausible relationships among the data. As a consistency check, we also checked for evidence of relationships amongst the raw $L_{X}$ values finding that the same correlations are still present in the data, albeit at lower significance level. The presence of the same correlations in both the contiguous $L_{X}$ and the non-simultaneous $\lbol$ indicates that the correlations are not driven by uncertainties in $\lcorr$.

\section{Discussion}
\label{sec:discussion}

\subsection{Comparisons with previous work}
\label{sub:comparisons_with_previous_work}

\subsubsection{The \xmm outflow sample}
\label{sub:the_xmm_outflow_sample_}
In Paper~I we showed that $\sim40\%$ of the AGN in the \suzaku sample exhibit evidence for highly-ionised winds in their Fe\,K band. This is in agreement with the detection fraction reported by \citet{tombesi2010a} for \xmm, and corroborates their conclusion that such winds are either (i) persistent over the active phase and thus have a large covering fraction, or (ii) transient in nature and have a covering fraction of unity for only a fraction of the time. We also showed that the $\lognh$ and $\logxi$ distributions are also entirely consistent between the two samples, covering ranges of $21\lesssim\log(\nh/{\rm cm}^{-2})\leq24$ and $2\lesssim\logxi\leq6$ in both, respectively, with the outflow velocity $\vout$ also being largely consistent on the basis of a Kolmogorov-Smirnov (KS) test. By comparing the distributions of wind parameters found in this work to those determined by \citet{tombesi2010a} we similarly find that the Fe\,K absorbers detected in both the \suzaku and \xmm studies are typified by the same range of physical parameters and occupy the same parameter space in terms of their overall location and energetics: $1\lesssim\log(r/\rs)\lesssim3$ ($\sim0.001-1$\,pc), $-2\lesssim\log(\mout/\medd)\lesssim0$ ($\sim0.001-10$\,\msun\,yr) and $-3\lesssim\log(\lk/\ledd)\lesssim0$. Their medians are also similar with $\log(r/\rs)\sim16(17)$, $\log(\mout/{\rm g\,s}^{-1})\sim25(25)$ and $\log(\lk/{\rm erg\,s}^{-1})\sim44(44)$ for \suzaku (\xmm), respectively, in natural units.

The \suzaku outflow sample therefore robustly confirms that Fe\,K absorption are a real physical component of emergent X-ray spectrum, that the implied winds likely have a large covering fraction, and that they are typified by large column densities and high ionisation. The range of outflow velocities, which are typically $\vout\geq10,000$\vunit, but can be as low as $\lesssim1,000$\vunit in some cases, also suggests that the highly-ionised winds share an overlap in velocity space with the traditional soft X-ray warm absorber. This is in line with the results of \citet{tombesi2012c} who argue that some of the soft X-ray warm absorbers could be the artefacts of accretion disc winds which have propagated farther away from the SMBH.

\subsubsection{Comparison with King et al. (2013)}
\label{sub:king2013}
Another interesting study that we can consider is that of \citet{aking2013}, who compare the kinetic power of both the warm absorber and relativistic jets to $\lbol$ of their respective source. They consider both black-hole binaries (BHBs) and AGN which means that their study samples an extremely broad spectrum of both black hole mass $[0.8\lesssim\log(\mbh/\msun)<10]$ and bolometric luminosity [$37\lesssim\log\lbol<47$]. Even though the \citet{aking2013} study deals primarily with the low-velocity soft X-ray warm absorber they do include a few `ultra-fast outflows' (which they define as those with $\vout\geq0.01$\,c, as we have done in this work) as a matter of comparison, although they do not fit them directly as part of their regression analysis; two of their `ultra-fast' sources, namely 3C\,111 and APM\,08279+5255, are also included in our \suzaku sample. In principle we could compare the properties of the \suzaku-detected outflows with those of King et al. to determine if our high velocity Fe\,K winds scale in a similar manner to the warm-absorber. However, this is not possible in practice because the two studies compute the wind kinetic power according to different assumptions (King et al. give their quantities in units of per-covering-fraction), which means the intercept of any linear regressions will therefore by intrinsically offset. Even so, the {\it slope} of any linear fit --- which is the important parameter as it measures how the kinetic power varies with bolometric luminosity --- can still be compared because it is largely independent of the normalisation. 

King et al. find the kinetic power of warm absorber to be strongly correlated with $\lbol$, scaling with a tight global slope of $\log\lk\propto(1.58\pm0.07)\log\lbol$, while they found that those in AGN have a much flatter local slope of $0.63\pm0.30$ when considered in isolation (although that this could be driven by there being a smaller number of AGN). In Figure~\ref{fig:lbol_raw}d we showed that kinetic power of the \suzaku-detected Fe\,K winds are also strongly correlated with the bolometric luminosity with a slope in the range $1.5^{+1.0}_{-0.8}$ ($R_{p}\geq0.54$). This result, which is driven by the $\vout\propto\lbol^{0.5}$ dependence, is in good agreement with the slope reported by King et al. for their global sample and suggests that the fast Fe\,K winds may scale with $\lbol$ in a similar manner to the warm absorber. To summarise, by comparing our results for those obtained by \citealt{aking2013} we see that the fast Fe\,K winds ($\vout>0.01$\,c) is similar to that found for slow warm absorbers.

\subsection{Are the winds radiatively accelerated?}
\label{sub:are_the_winds_radiatively_accelerated_}
In this work we shown that the observed wind velocity, and subsequently the overall wind energetics, are proportional to the AGN bolometric luminosity. It is therefore tempting to conclude that the winds are accelerated by radiation pressure. In this section we investigate whether radiation pressure alone can account for the observed correlations. In contrast to the undoubtedly line-driven winds which are prevalent in the UV spectra of AGN, the high ionisation state of the Fe\,K winds suggests that line-driving is unlikely to be the dominant acceleration mechanism here. An alternative means of radiatively accelerating high ionisation gas is through Thomson/Compton scattering of the continuum X-ray photons. While line-driving may play a role in large black holes accreting near the Eddington limit (Hagino et al. 2014, in prep) we concentrate on the latter mechanism here.

 In comparison to fast line-driven winds in the UV, the comparatively small interaction cross-section in highly ionised material means that high-ionisation winds, such as those considered here, couple much less efficiently with the incident radiation and, as a result, require either much larger source luminosities or higher column densities to achieve equivalent outflow velocities. A general characteristic of these `continuum-driven winds' is that they are accelerated by photons scattering off free electrons in the absorbing gas which transfers a portion of the photon momentum to the material, hence causing a wind. This scenario has been considered extensively in the literature (e.g., \citealt{king2003, king2005, king2010,reynolds2012,costa2014}). We outline the general theory again here in an effort to search for relevant relationships with $\lbol$ which we can relate to the results of our regression analysis.

\begin{figure*} 
	\begin{center}
		\includegraphics[width=0.45\textwidth]{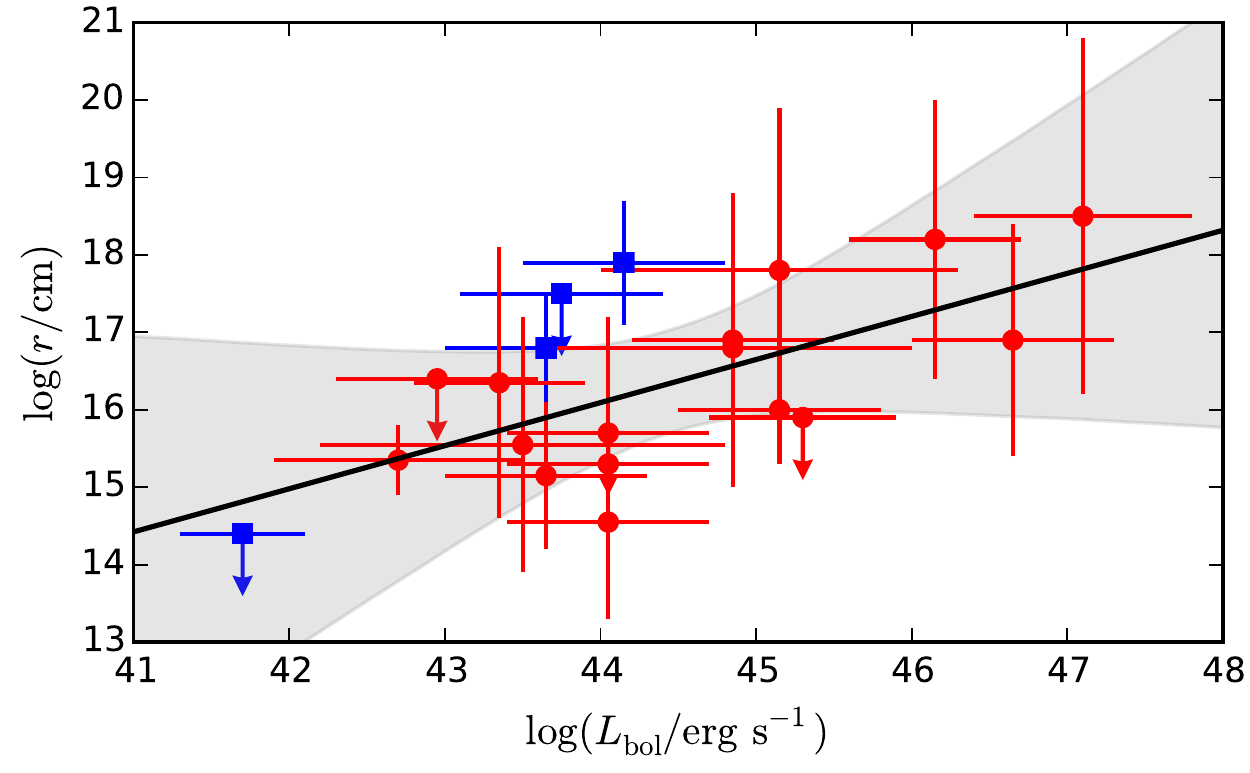}
		\includegraphics[width=0.45\textwidth]{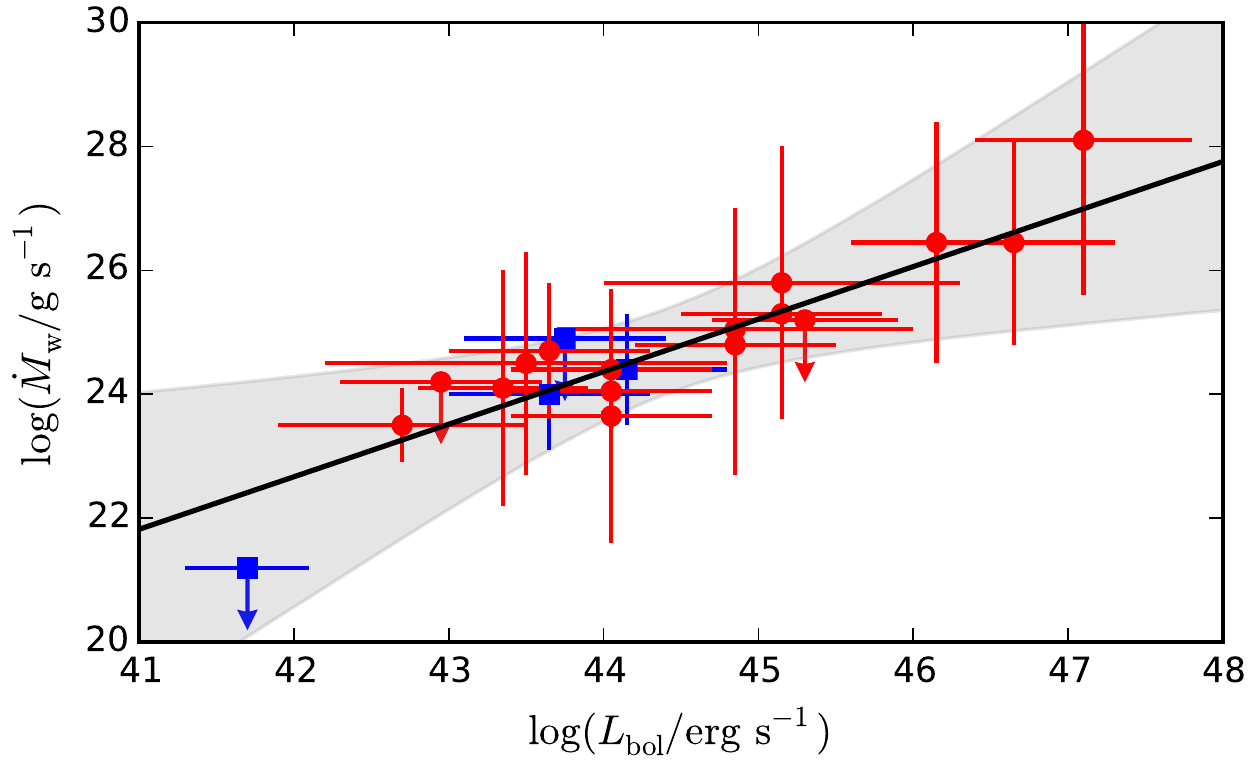}
		\includegraphics[width=0.45\textwidth]{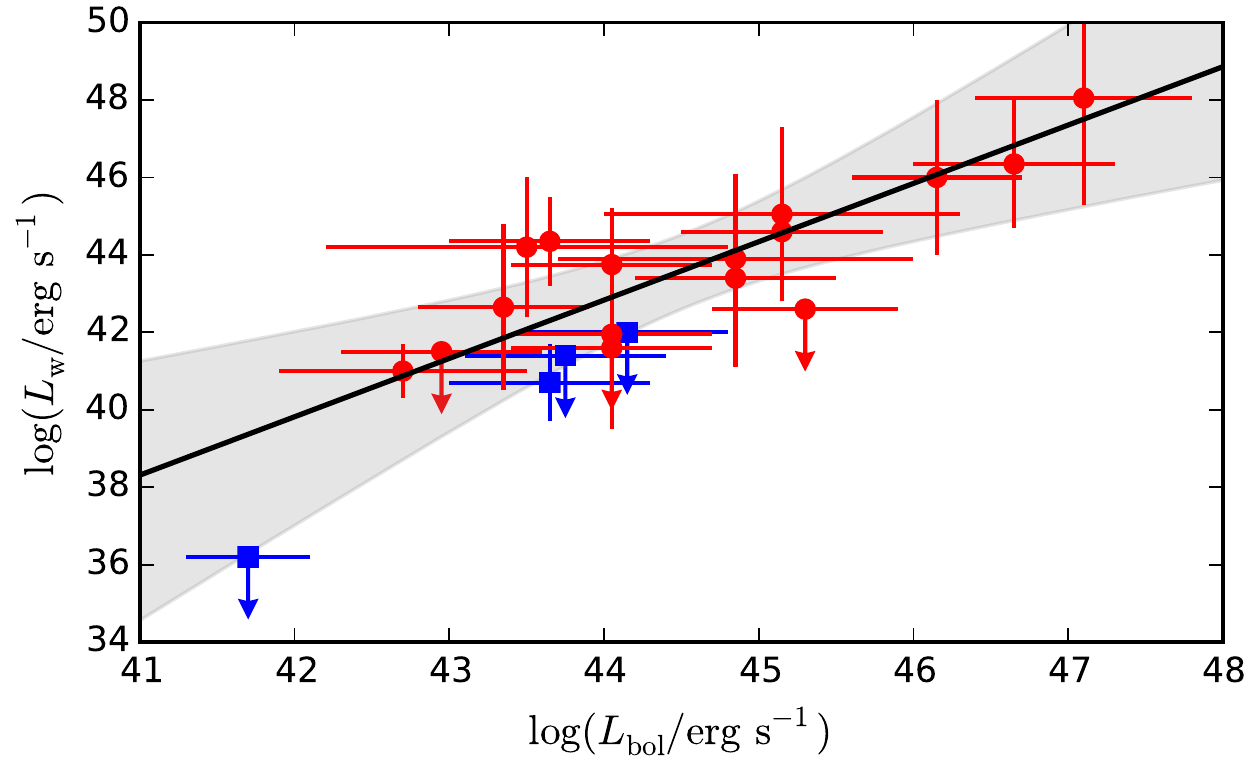}
		\includegraphics[width=0.45\textwidth]{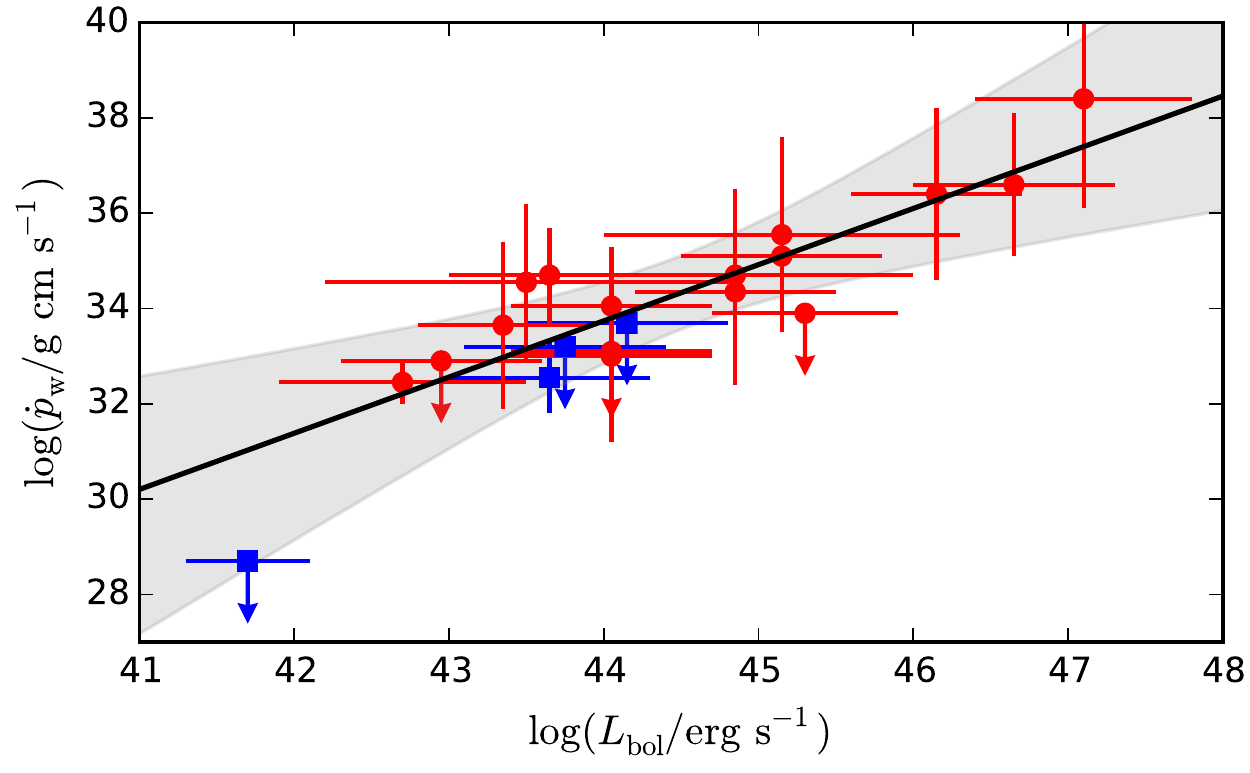}
	\end{center}
	\caption{Scatter plots showing $\log\lbol$ against the raw wind parameters: (a) $\log(r/{\rm cm})$, (b) $\log(\mout/{\rm g\,s}^{-1})$, (c) $\log(\lk/{\rm erg\,s}^{-1})$ and (d) $\log(\pout/{\rm g\,cm\,s}^{-1})$. Data points and plot components have the same meaning as in previous figures.}
	\label{fig:lbol_raw}
\end{figure*}

The efficiency through which the photon momentum of the incident radiation field is transferred into the wind depends upon the plasmas optical depth to electron scattering, $\tau$, which determines how many times each electron is likely to scatter in the flow. The total electron scattering optical depth of a homogeneous wind viewed from infinity down to radius $r$ is given by
\begin{equation}
	\tau=\nh\sigma_{\rm T}=\sigma_{\rm T}\int_{r}^{\infty}n \cdot dr = \dfrac{\sigma_{\rm T}\mout}{4\pi b m_{\rm p} \vout r},
	\label{eqn:tau_1}
\end{equation}
where $\sigma_{\rm T}$ is the Thomson cross-section for electron-scattering, $n=\mout/4\pi b m_{\rm p} \vout r^{2}$ is the electron number density in a mass-conserving spherical flow (c.f., \S\ref{sub:mass_outflow_rate}) and $b$ is a geometrical factor which takes into account the global covering of the gas. Normalising to the Eddington mass-accretion rate, $\medd=\ledd/\eta c^{2}=4 \pi G \mbh m_{\rm p}/\sigma_{\rm T}\eta c$, then yields
\begin{equation}
	\tau=\dfrac{1}{b \eta c}\dfrac{1}{\vout}\dfrac{G\mbh}{R}\dfrac{\mout}{\medd},
	\label{eqn:tau_2}
\end{equation}
as has been shown by \cite{king2003} (see also \citealt{king2010} for a recent review). If we further assume that the measured flow velocity roughly corresponds to the escape velocity at the radius where the wind was launched, i.e., $R=R_{\rm esc}=2G\mbh/v^{2}$, we have
\begin{equation}
	\tesc=\dfrac{1}{2b\eta}\dfrac{\vout}{c}\dfrac{\mout}{\medd},
	\label{eqn:tau_3}
\end{equation}
which, after substituting for $\medd=\ledd/\eta c^{2}$ and re-arranging, becomes
\begin{equation}
	\pout={2b}\tesc\dfrac{\ledd}{c}=2b\dfrac{\tesc}{\eddratio}\dfrac{\lbol}{c}%\simeq\dfrac{\tesc\ledd}{c}\simeq\dfrac{\tesc}{\eddratio}\dfrac{\lbol}{c},
	\label{eqn:pout_pbol}
\end{equation}
where $\pout\equiv\mout\vout$, the geometric factor $b=\Omega/4\pi$ and we have substituted from $\eddratio = \lbol/\ledd$ in the final step to express the relation in terms of observable quantities. Equation~(\ref{eqn:pout_pbol}) therefore shows that a wind accelerated by electron scattering should have an outward momentum-rate which is proportional to that of the incident radiation field. This result (and the associated derivation) is directly equivalent to the one presented by \cite{king2003}. Note that while $\tesc$ has thus far been described as the optical depth to electron scattering it can in principal also account for additional sources of opacity in the flow, e.g., though bound-free or bound-bound absorption, which can also be further boosted by multiple scattering events. A reasonable way of testing for the continuum-driven wind scenario is to therefore see whether $\pout$ and $\pbol\equiv\lbol/c$ follow a linear trend. Figure~\ref{fig:pbol_pout}a shows that the two are clearly correlated, with a slope of $\beta=1.2^{+0.8}_{-0.7}$. This is consistent with the order of unity expected from equation~(\ref{eqn:pout_pbol}), and is also agreement with the slope of $\beta_{\rm xmm}\sim1.6\pm1.1$ found by \citet{tombesi2012c}. Figure~\ref{fig:pbol_pout}b is a constraint diagram for $\log(\pout/\pbol)$ showing the deviations from the expected ratio of unity; in total, most of the sources (13/20; $65\%$ of the sample) are formally consistent with the ratio of unity expected for a continuum-driving scenario. 

Another useful relationship can be obtained by multiplying equation~(\ref{eqn:pout_pbol}) again by $\vout$:
\begin{equation}
	\lk=b\tesc\dfrac{\vout}{c}\ledd=b\dfrac{\tesc}{\lambda}\dfrac{\vout}{c}\lbol,
	\label{eqn:lk_lbol}
\end{equation}
i.e., the wind kinetic power is predicted to be proportional to the bolometric luminosity of the AGN. In Figure~\ref{fig:lbol_raw}d we showed that $\lk$ and $\lbol$ are indeed correlated (Table~\ref{tab:linear_regression}), although the slope ($\beta\sim1.5^{+1.0}_{-0.8}$) appears to be slightly steeper than (but still marginally consistent with) that predicted from equation~(\ref{eqn:lk_lbol}). The steeper slope here is probably due to the $\vout\sim\lbol^{0.5}$ relationship observed before. Indeed, taking this into account yields $\lk\propto\lbol^{0.5}\times\lbol=\lbol^{1.5}$, which is exactly what we observe in these data. 

Even so, it is worth noting that other acceleration mechanisms are not conclusively ruled out for these winds. If the winds are launched with zero initial radial velocity, then the terminal velocity of the wind in terms of the Eddington limit is: $v_{\rm inf}=\sqrt{2G\mbh/R_{\rm launch}(f_{\tau}\lbol/\ledd-1)}$, where $v_{\inf}$ is the local terminal velocity and $f_{\tau}$ is the force multiplier ($\equiv1$ for Thomson scattering). As a result, any wind accelerated purely by Thomson-scattering will not be able to greatly exceed the local terminal velocity without either: (a) additional sources of opacity in the flow (such that $f_{\tau}>1$), or (b) additional mechanisms transferring momenta to the flow. In this case, additional opacity in the flow may be attributed to the outflowing gas if it is stratified and contains clumps of lower-ionisation gas. This extra opacity could perhaps be associated with the partially-covering gas required by many of our models in Paper~1. Alternatively, magnetic processes may play a role in the initial acceleration which provides additional momentum to the flow, thereby allowing it to escape the system. Indeed, have been proposed along these lines in the literature (e.g., \citealt{ohsuga2009,ohsuga2012,ohsuga2014}. It is not possible to distinguish between these possibilities on the basis of existing data.

\subsubsection{Estimating the wind opacity}
\label{sub:estimating_tau}
\begin{figure}
	\begin{center}
		\includegraphics[width=0.4\textwidth]{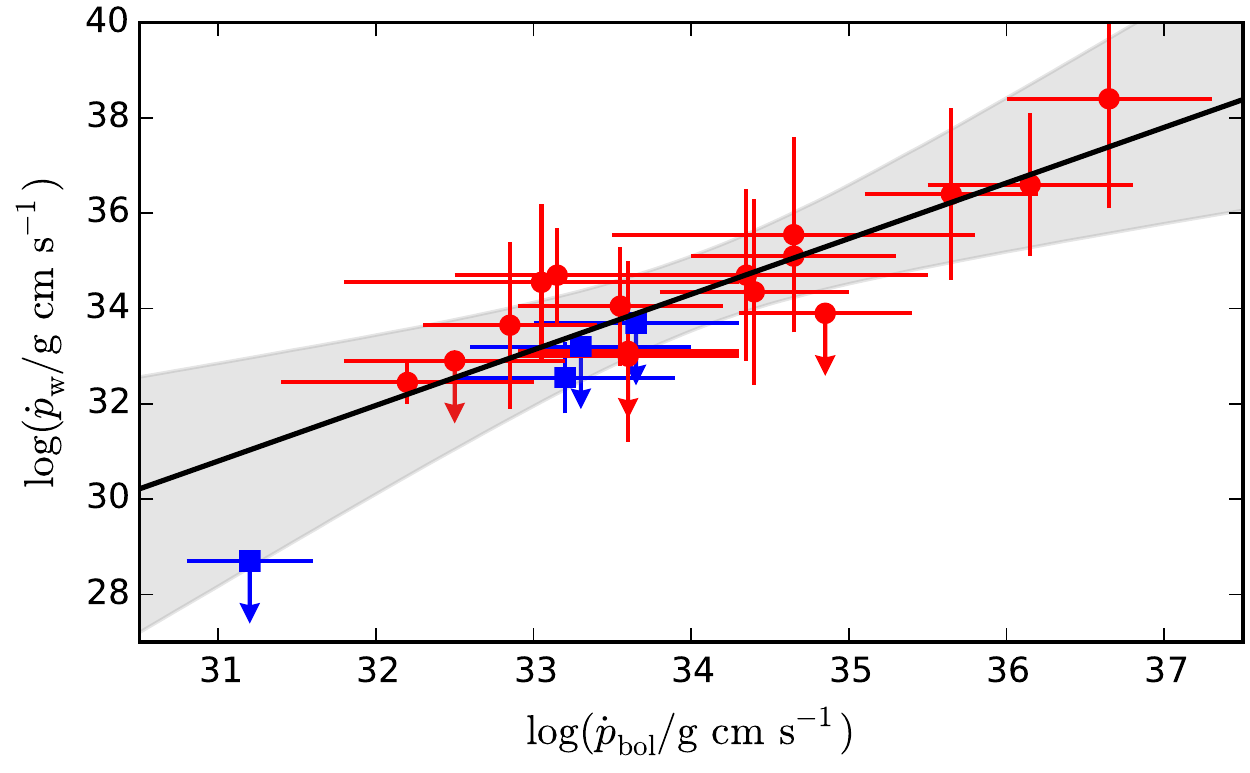}\vspace{-0.73cm}
		\includegraphics[width=0.4\textwidth]{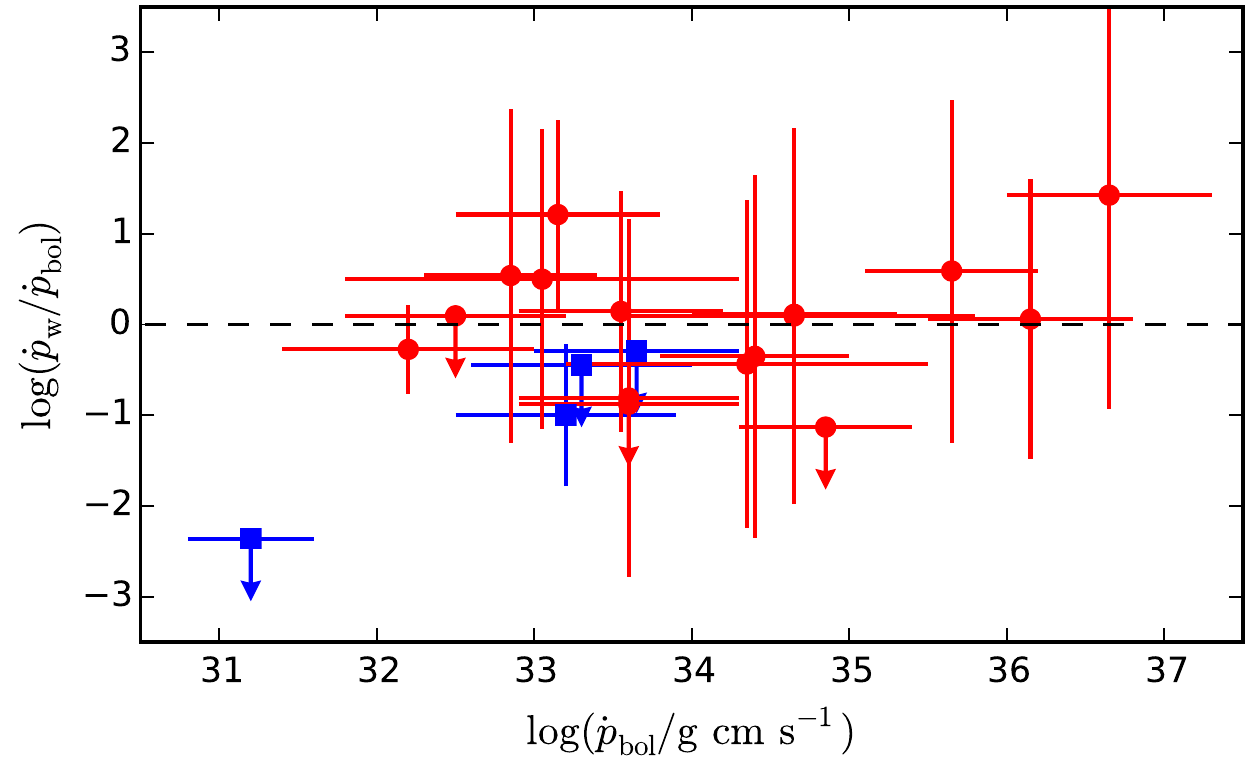}
	\end{center}
	\caption{Comparison of the outflow ($\pout$) and the radiation field ($\pbol$) momentum rates. {\it Top}: a scatter plot showing the clear linear relationship between the two quantities (see Table~\ref{tab:linear_regression} for parameters). Plot components are the same as previous. {\it Bottom}: Constraint diagram showing the $\pout/\pbol$ ratio for each source. Almost all of the AGN in the sample are consistent with a ratio of unity (denoted by the dotted black line). }
	\label{fig:pbol_pout}
\end{figure}

\begin{figure}
	\begin{center}
		\includegraphics[width=0.4\textwidth]{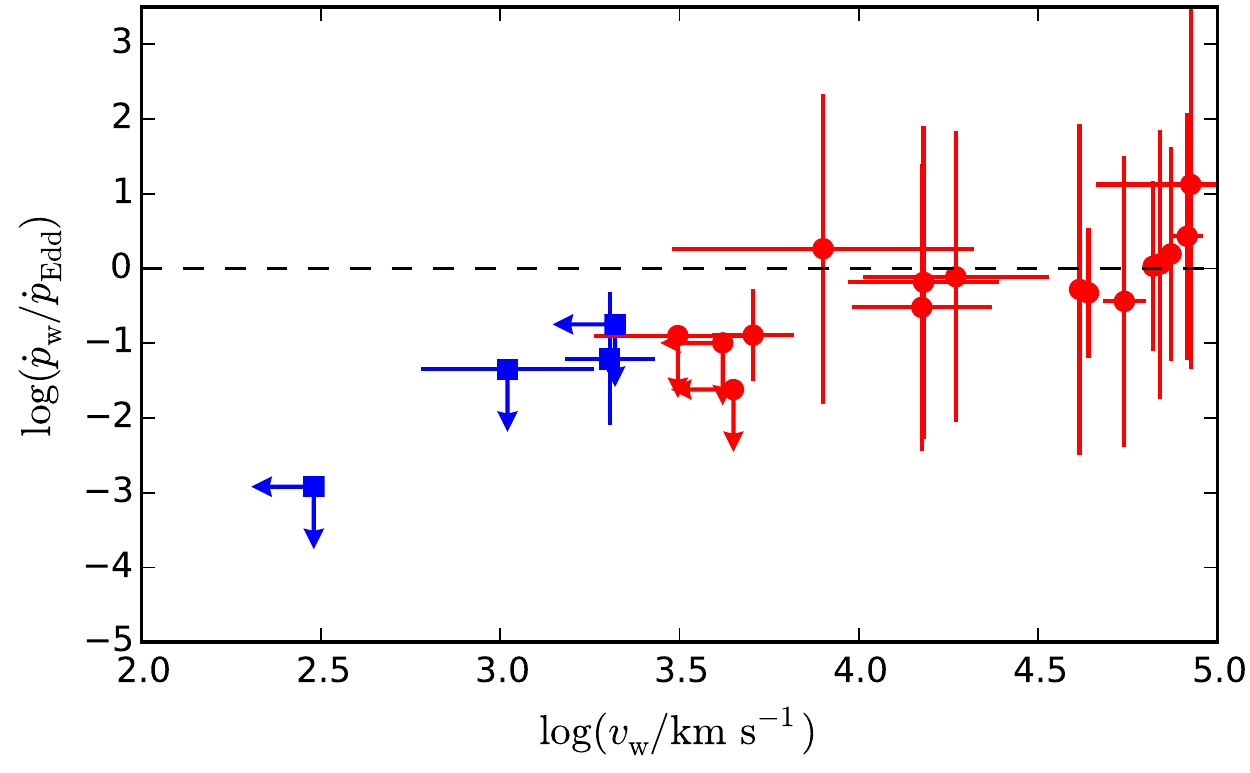}
	\end{center}
	\caption{Plot showing $\log(\pout/\pedd)$ versus $\log(\vout/{\rm km\,s^{-1}})$ which can be used as a diagnostic for $\tau$ (see text for details). The majority of fast AGN ($\vout>3000$\,km\,s$^{-1}$; red points) are consistent with $\pout=\pedd$, thereby implying that $\tau=1$, while all of the slow winds (blue points) are well below a ratio of unity.}
	\label{fig:pout_tau}
\end{figure}

Considering again the continuum-driven scenario, an interesting thing of note from Figure~\ref{fig:pbol_pout}b is that while all of the fast Fe\,K winds are consistent with the unity ratio within the errors, all of the slow winds fall below the limit (even when only the upper-limits are considered). This possibly reflects subtle changes in the $\tesc/\eddratio$ ratio which allows for the possibility of continuum-driven winds in sub-Eddginton AGN, whilst also allowing for subtle differences in the coupling efficiency of the gas between the two groups. As mentioned in \S\ref{table:measured_wind_properties}, the AGN considered here are relatively tightly clustered around a mean $\eddratio\sim0.1$ which means that we are unable to reliably test how the Eddington ratio affects the $\pout/\pbol$ ratio. Nevertheless, we can still try to gauge how $\tau$ changes across the sample. From equation~(\ref{eqn:pout_pbol}) we can see that when normalised to Eddington $\pout/\pedd=2b\tau$, with $b\sim \Omega/4\pi \approx 0.4$ (see \S\ref{sub:mass_outflow_rate}) and therefore that $2b\simeq1$ to within an order of magnitude. This means that $\pout/\pedd\simeq\tau$, and that the $\pout/\pedd$ ratio can be in principal be used as a proxy for the gas opacity when the wind was launched. Figure~\ref{fig:pout_tau} shows how $\log(\pout/\pedd)$ varies with $\vout$. The higher velocity systems are largely consistent with $\pout=\pedd$ ($\tau=1$) within the errors, with a mean of $\langle\tau\rangle_{\rm fast}\sim0.3$, which is consistent with what is argued by \citet{king2003}. In contrast, the slow systems (blue) all appear to have $\pout/\pedd$ ratios around an order of magnitude lower. Taken at face value, this possibly indicates that they have a lower opacity to scattering. There also appears to be a transitionary phase between the two groups, with $\vout\sim$ a few thousand $km\,s^{-1}$, which may be consistent to the idea of stratified winds launched over a wide range of radii with varying velocities (see \citealt{tombesi2012c}). It is worth noting, however, that the opacity estimated from $\pout/\pedd$ could be lower than that of the bulk flow if (a) we view the wind from an acute angle where the observed velocity is not representative of the bulk flow velocity, and (b) the wind has a component of completely ionised gas which contributes to the scattering opacity but has no spectroscopic signatures. One possibility is that all of the winds have $\tau\simeq1$ when integrated over all lines of sight and across all ionisation phases. Alternatively, the fact that the slow winds have $\tau\ll1$, coupled with the fact that they are preferentially observed in lower luminosity AGN, may indicate that they are accelerated by another mechanism such as magnetic pressure, for example, which may have a larger contribution at lower luminosities (e.g., see \citealt{ohsuga2009,ohsuga2012}) 

Regardless, the fast winds --- which are the ones most likely to affect their host galaxy due to their large velocity and therefore the most interesting in terms of AGN--host-galaxy feedback --- are consistent with a wind launched by continuum scattering at the Eddington limit (e.g., \citealt{king2003}).

\subsection{Energetic significance and feedback implications}
\label{subsec:energetic_significance}
\begin{figure}
	\begin{center}
		\includegraphics[width=0.4\textwidth]{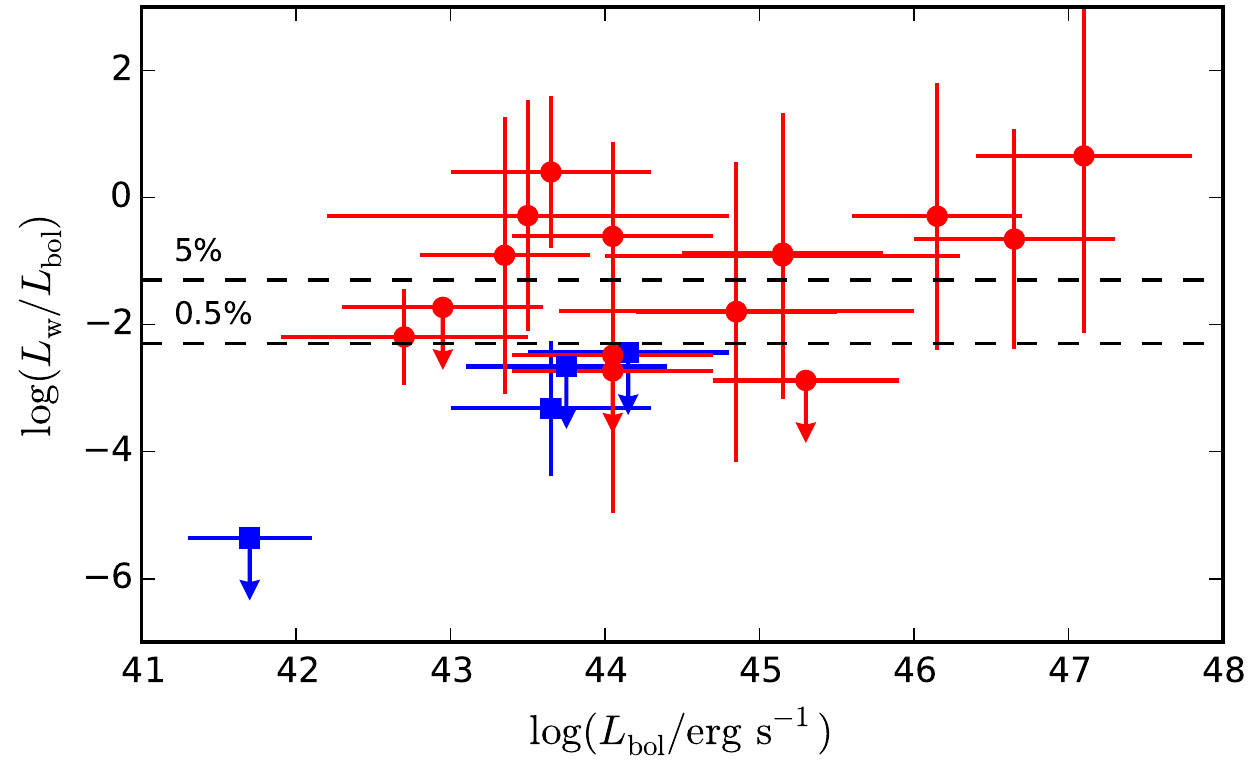}
	\end{center}
	\caption{Diagnostic plot comparing the ratio of $\lk/\lbol$ to the $\lbol$ of each AGN in the \suzaku sample. The horizontal lines denote $\lk/\lbol$ fractions of $0.5\%$ and $5\%$ which are the theoretical thresholds for feedback. Almost all of the fast absorption systems (red points) have $\lk$ in excess of $0.5\%$ of $\lbol$ which suggests that they may be significant in terms of AGN--host-galaxy feedback scenarios.}
	\label{fig:lbol_kout}
\end{figure}

The independent \suzaku and \xmm wind samples have conclusively shown highly ionised Fe\,K winds are both frequently observed in AGN and that they contain a large amount of mechanical energy. However, a fundamental question still remains: is the energy imparted by the mass flow likely to play a role in terms of AGN--host-galaxy feedback scenarios? AGN feedback models have postulated that $\sim5\%$ of an AGNs' bolometric radiative output needs to be converted to mechanical power in order to have a notable effect on the growth of a central SMBH and the host galaxy (e.g., \citealt{dimatteo2005}). However, recent numerical simulations by \citealt{hopkins2010} have shown that should the effects of `secondary' feedback such as cloud ablation be taken into account, the required energy can be significantly lower, of the order $\sim0.5\%$. Moreover, hydrodynamic simulations have also shown that even relatively modest Fe\,K outflows (e.g., those with only $\vout\sim0.01$\,c, $\mout\sim0.01$\msun\,yr$^{-1}$ and $\lk\sim10^{44}$\,erg\,s$^{-1}$; see \citealt{wagner2013}) can impart significant feedback upon the host galaxy on $\sim$kpc scales should these effects be taken into account. Investigating what fraction of an AGNs' bolometric output is conveyed through mechanical power therefore provides a means of qualitatively assessing the likely energetic significance of the mass flow. A plot of $\lbol-(\lk/\lbol)$ to this effect is shown in Figure~\ref{fig:lbol_kout}.

A large fraction of the outflows have a kinetic luminosity either in excess of or consistent with the $0.5\%$ of $\lbol$, with 10/20 also exceeding $5\%$ as well. The mean kinetic power considering only the fast winds is $\langle \lk/\lbol\rangle\approx7\%$. This is well above the supposed $\sim0.5\%$ threshold for significant feedback. In the most conservative case, i.e., by considering only $\lkmin$ given the available errors, we estimate $\langle \lk/\lbol\rangle\approx0.2\%$. This is comparable to the $\sim0.5\%$ threshold and consistent with what was found by \citet{tombesi2012a} on the basis of the \xmm outflow sample. Therefore, even in the most conservative case, the energetic output of the fast \suzaku-detected Fe\,K winds is comparable to the threshold thought necessary for feedback. Importantly, these estimates are derived solely from the wind kinematics and are therefore independent of the initial driving mechanism.

\subsection{Relation to $\msigma$ in quiescent galaxies}
\label{sec:relation_to_msigma_in_quiescent_galaxies}
The FeK winds detected here may therefore impart enough energy into the host galaxy to play a roll in galaxy-scale feedback. With this in mind, it is useful to highlight the recent work of \citet{mcquillin2013}. To set the context of their work, it is first useful to explain the difference between momentum- and energy-driven outflows. As a wind propagates into the host galaxy it will sweep up a shell of ambient gas, the dynamics of which are determined by how efficiently the shocked wind material behind the shell can cool. Initially, the shocked gas can cool efficiently (by Compton scattering) so the shell stays thin and is driven outwards directly by the force of the wind (cf. our equation~\ref{eqn:pout_pbol}; also \citet{king2003,king2010,mcquillin2012}). This is momentum-driven feedback. As the outflow reaches larger distances, however, the cooling efficiency drops and the shocked gas becomes geometrically thick and hot. The outwards force on the shell then comes from the wind thrust in equation (\ref{eqn:pout_pbol}) mediated by the thermal ppressure and work done on and by the shocked gas. This is energy-driven feedback (e.g., \citealt{king2003b,king2005,zubovas2012b,mcquillin2013}). The transition between the two phases can occur at relatively small radii (see \citealt{mcquillin2013}), and energy-driven outflows are therefore expected to be dominant on large galactic scales.

\citealt{mcquillin2013} consider purely energy-driven outflows and investigate their link to the observed $\msigma$ relationship. They showed that, for an energy-driven shell to reach the escape speed from an isothermal sphere of dark matter and gas with velocity dispersion $\sigma$ requires a critical BH mass given by
\begin{equation}
	\left(\dfrac{\vout}{c}\right)\left(\dfrac{\mbh}{10^{8}\msun}\right)=6.68\times10^{-2}\left(\dfrac{\sigma_{\ast}}{200\vunit}\right)^{5},
\end{equation} 
such that the velocity of the wind from the black hole ($\vout$) enters explicitly. McQuillin \& McLaughlin used this equation to infer the wind speed that would have been needed during the active protogalactic phase of 51 quiescent ($z\approx 0$) galaxy spheroids (taken from the sample of \citet{gultekin2009}) in order to explain the observed values of $M_{\rm BH}$ and $\sigma$ in these galaxies as a result of (energy-driven) feedback. Their main result is a distribution of BH wind speeds, derived in essence from the scatter of the $\msigma$ relation, that bears a remarkable resemblance to the empirical distributions of $\vout$ in the AGN samples of both \citealt{tombesi2010a} and Paper~I.

\begin{figure}
	\begin{center}
		\includegraphics[width=0.4\textwidth]{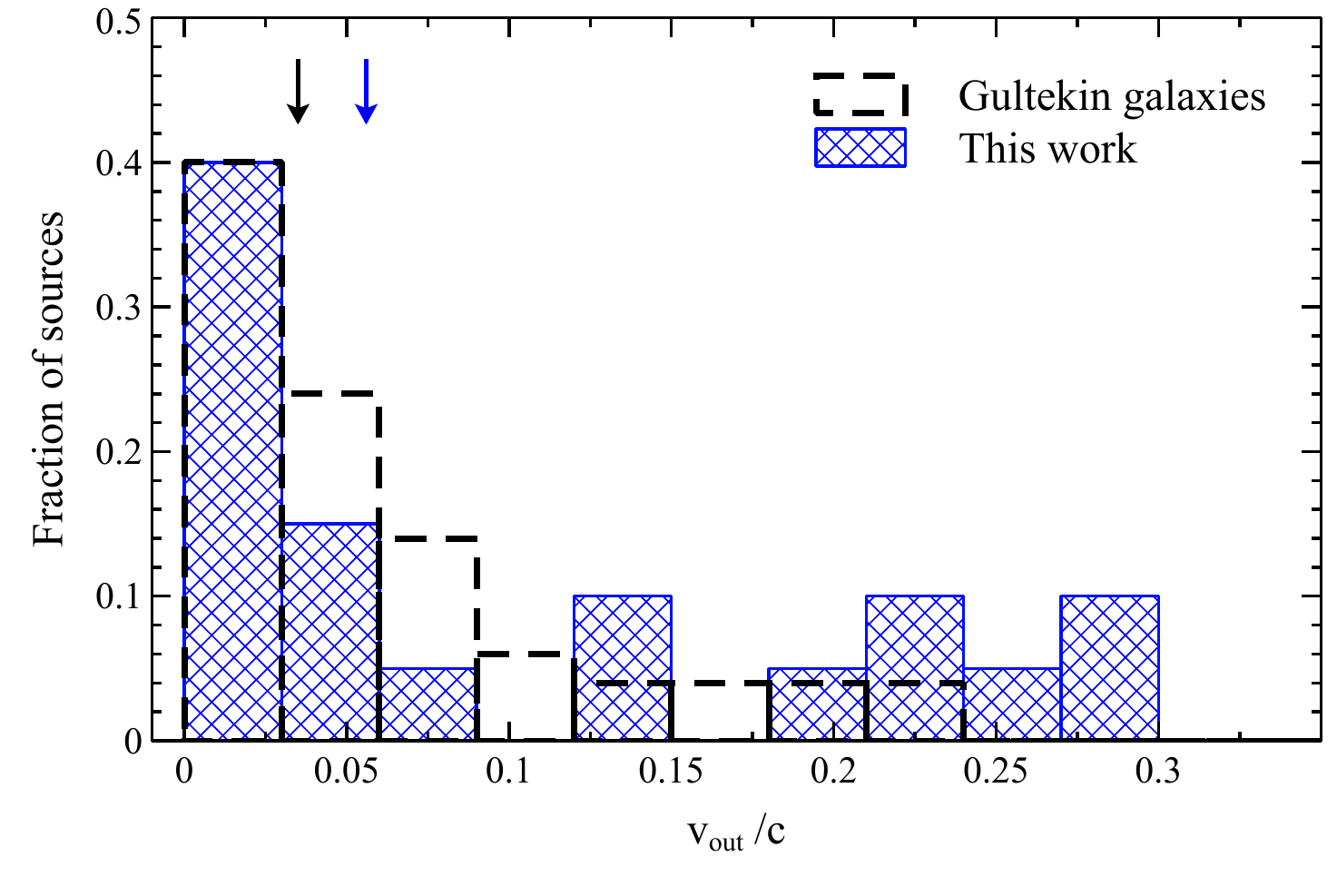} 
	\end{center}
	\caption{Histogram comparing the \suzaku-observed distribution of outflow velocities (blue cross-hatched area) to those inferred from local quiescent galaxies by \citet{mcquillin2013} who assume that their observed $\msigma$ relation (c.f., \citealt{gultekin2009}) is induced by energy-driven feedback, with $\mbh\vout\propto\sigma_{\ast}^{5}$. The blue and black arrows correspond to the median outflow velocites of $\vout\sim0.056$\,c and $0.035$\,c in the \suzaku-observed AGN and quiescent sub-samples, respectively. The two samples are formally indistinguishable by the Kolmogorov-Smirnov (KS) test. See text for further discussion.}
	\label{fig:mcquillin}
\end{figure}

Figure~\ref{fig:mcquillin} compares the velocity distribution inferred by \citet{mcquillin2013} to the one that we measured for \suzaku-detected Fe\,K outflows (note that we consider all of the outflows in this panel). As noted by McQuillin \& McLaughlin, there is a close resemblance between the two, with both samples tracing a similar overall distribution. Their overall range are similar, as is their median: $\vout\sim0.035$\,c in the quiescent galaxies, and $\sim0.056$\,c for the \suzaku-detected outflows (see Paper I). Formally, the two distributions cannot be distinguished: a Kolmogorov-Smirnov (KS) test only rules their being from the same parent population at $P_{\rm KS}\sim75\%$ confidence. The fact that there are similarities between the inferred and observed $\vout$ distributions at all is in itself extremely interesting, not only because the independent samples of quiescent galaxies and AGN are drawn from two different sub-populations of galaxy, but because it implies a close relationship between the emergent wind and the evolution of the host galaxy. Overall, the work of \citet{mcquillin2013} is encouraging for the general notion that high-velocity outflows contribute to the feedback which is invoked to explain the $\msigma$ relationship in local quiescent galaxies, and, more importantly, for the idea that AGN play a key role in galaxy formation and evolution.

\section{Summary \& conclusions}
\label{sec:summary_conclusions}
In Paper~I we formed a detailed spectroscopic study of Fe\,K absoption in a heterogeneous sample of 52 \suzaku-observed AGN. There, we found that $\sim40\%$ of the sources in the sample harboured statistically significant \fexxvi~\hea and/or \fexxvi~\lya absorption lines with velocities ranging from a few thousand km\,s$^{-1}$ to $\sim0.3$\,c. We also measured their column density ionisation parameter of the absorber using the \xstar photoionisation code. Here, we build upon these results to compute the location $r$, mass outflow rate $\mout$, kinetic power $\lk$ and momentum flux $\pout$ of the implied wind, and use these results to inform an exploratory analysis of how the wind parameters scale with the bolometric luminosity of their host AGN. We focus our attention on the fastest winds, defined here as $\vout>0.01\,c$, because these ones likely represent genuine disk-winds which are likely distinct from the slower (and more distant) warm absorber. The main results of our study are summarised below.
\begin{enumerate}[label={(\roman*)},leftmargin=*]
	\item The mean radial distance to the fast winds is $\langle r \rangle\sim10^{15-17}$\,cm (typically $\sim10^{2-4}\,\rs$) from the black hole. This corresponds to $\sim0.0003-0.03$\,pc, such that the Fe\,K absorbers are located much closer to the black hole than the traditional parsec-scale warm absorber. The mean wind mass outflow rate and mean kinetic power are constrained to $\langle \mout \rangle\sim10^{24-26}$\,g\,s$^{-1}$ ($\sim0.01-1\,\mout\,{\rm yr}^{-1}$) and $\langle \lk \rangle\sim10^{43-44}$\,erg\,s$^{-1}$, respectively. The average upper limit is $\sim\medd$, but is still an appreciable $\sim1\%$ of $\medd$ if only the lower limits are considered, whereas the kinetic power is constrained to $\sim(0.01-0.1)\ledd$. These properties are consistent with wind at a substantial fraction of the Eddington limit which carries a large amount of mass into the host galaxy.
	\\
	
	\item The Fe\,K absorber column density and ionisation parameter are largely independent of $\lbol$ but there is a significant correlation between the wind velocity $\vout$ and $\lbol$, with a slope $\beta=0.4^{+0.3}_{-0.2}$. This tentatively indicates that more luminous AGN launch winds with a larger observed velocity along the line-of-sight.
	\\
	\item The winds in more luminous AGN contain more material, and they are subsequently more energetic. The median slope of the $\lbol-\lk$ and $\lbol-\pout$ regressions are formally consistent with unity.
	\\
	\item There are no significant correlations in the $\lbol-(r/\rs)$ and $\lbol-(\mout/\medd)$ planes. Therefore, the winds are located at similar distances in Schwarzschild units and contain a similar fraction of $\medd$.
	\\
	
	\item The momentum flux of the radiation field ($\pbol$) and that of the wind ($\pout$) are strongly correlated, with a slope $\beta=1.2^{+0.8}_{-0.7}$. This is quantitatively consistent with what is expected from a continuum-driven wind.
	\\

	\item A significant fraction of the sample ($17/20$, $85\%$) exceed the minimum $\lk/\lbol\sim0.5\%$ threshold thought necessary for feedback (\citealt{hopkins2010}), while $9/20$ ($45\%$) also exceed the less conservative $\sim5\%$ threshold as well (\citealt{dimatteo2005}). In the most conservative case, the mean $\lk/\lbol$ ratio is $\sim0.2\%$. This suggests that the winds may be sufficiently energetic in terms of feedback. 
\end{enumerate} 

\noindent These results enforce those recently obtained with \xmm and provide additional evidence in favour of Fe\,K absorption being a prevalent feature of the AGN X-ray spectrum, and that the ensuing wind may be energetically significant in terms of feedback.

\section*{Acknowledgements} 
We thank the referee for their thorough reading of the manuscript and useful comments which greatly enhanced the clarity of this analysis. M.~Cappi acknowledges support from contracts ASI/INAF n.I/037/12/0 and PRIN INAF 2012. F.~ Tombesi acknowledges support for this work by the National Aeronautics and Space Administration (NASA) under Grant No. NNX12AH40G issued through the Astrophysics Data Analysis Program, part of the ROSES 2010.

\footnotesize{
	\bibliographystyle{mn2e_new}
	\bibliography{paper2}
}

\end{document}